\documentclass[a4paper,fleqn]{cas-sc}
\usepackage[numbers,square]{natbib}

\usepackage[T1]{fontenc}
\usepackage[utf8]{inputenc}
\usepackage{amsmath,amssymb}
\usepackage{graphicx}
\usepackage{booktabs}
\usepackage{subcaption}
\usepackage{xcolor}
\usepackage{hyperref}
\usepackage[section]{placeins}
\usepackage{float}
\usepackage{needspace}
\usepackage[most]{tcolorbox}

\graphicspath{{./}}

\ExplSyntaxOn
\bool_gset_true:N \g_stm_nologo_bool
\cs_set:Npn \__first_footerline:
  {
    \group_begin:
    \small
    \sffamily \__short_authors: :~
    { \rmfamily \itshape Preprint }
    \group_end:
  }
\ExplSyntaxOff

\newcommand{\E}{\mathbb{E}}
\newcommand{\Var}{\mathrm{Var}}
\newcommand{\sigmoid}{\sigma}
\newcommand{\I}{\mathbb{I}}
\newcommand{\NLL}{\mathrm{NLL}}
\newcommand{\NLLnorm}{\mathrm{NLL}_{\mathrm{norm}}}
\newcommand{\Zpennorm}{\mathrm{Zpen}_{\mathrm{norm}}}

\makeatletter

\let\casorigsection\section
\let\casorigsubsection\subsection
\renewcommand{\section}{\Needspace{7\baselineskip}\casorigsection}
\renewcommand{\subsection}{\Needspace{5\baselineskip}\casorigsubsection}
\makeatother

\begin{document}
\let\WriteBookmarks\relax
\setcounter{topnumber}{5}
\setcounter{bottomnumber}{5}
\setcounter{totalnumber}{8}
\renewcommand{\topfraction}{0.95}
\renewcommand{\bottomfraction}{0.85}
\renewcommand{\textfraction}{0.05}
\renewcommand{\floatpagefraction}{0.8}

\shorttitle{}    

\shortauthors{A.M. and P.M.}  

\title [mode = title]{Penalized Maximum Likelihood Inference of Core--Periphery Networks}


\author[1]{Antonio Mosca}[orcid=0009-0002-4023-4664]
\cormark[1]
\ead{a.mosca7@studenti.unipi.it}
\ead{antonio.mosca@unive.it}
\affiliation[1]{organization={Dipartimento di Fisica, Universit\`a di Pisa},
            addressline={Largo Bruno Pontecorvo 3}, 
            city={Pisa},
            postcode={56127}, 
            state={Tuscany},
            country={Italy}}

\author[2]{Piero Mazzarisi}[orcid=0000-0002-7198-513X]
\ead{piero.mazzarisi@unisi.it}
\affiliation[2]{organization={Dipartimento di Economia Politica e Statistica, Universit\`a di Siena},
            addressline={P.zza S.Francesco, 7-8}, 
            city={Siena},
            postcode={53100}, 
            state={Tuscany},
            country={Italy}}

\cortext[1]{Corresponding author}

\date{\today}

\begin{abstract}
Likelihood-based network models are often fitted under links' independence and low-order constraints, while empirical networks frequently exhibit systematic higher-order structures such as triangles and wedges, characterizing the observed clustering patterns. Real-world core--periphery networks such as the interbank market or the air transportation system represent key examples, with cores displaying complex and nonlinear features. We formalize \emph{Penalized Likelihood with Structural Discrepancies} (PLSD) as an inference-level correction that trades likelihood fit for agreement with targeted motifs. PLSD augments the negative log-likelihood with a penalty on standardized wedge and triangle discrepancies, yielding a controlled distortion of the likelihood surface that can be interpreted through a linear response analysis. A calibration approach for the unique tuning hyperparameter is introduced to turn off the penalization term when the model is correctly specified, thereby recovering maximum-likelihood estimation in that case.
We then introduce a mixed-constraint maximum-entropy core--periphery Exponential Random Graph Model (ERGM), derive unconditional semi-closed motif formulas under Pareto-distributed core fitness, and interpret sparse-regime scaling through a big-jump mechanism for heavy-tailed distributions. We finally corroborate the PLSD inference methodology both with Monte Carlo simulations of a toy stochastic block model and by applying the novel core--periphery ERGM to weekly eMID interbank networks (2009--2015) and monthly US air traffic networks (1991--2000). We show that PLSD describes cores that are not only dense but also higher-order-rich and captures the observed heterogeneous degree distributions and the overexpression of wedges and triangles, at a link-level cost proportional to the model misspecification.
\end{abstract}

\begin{keywords}
Network inference \sep Exponential Random Graph Models \sep Core--periphery networks \sep Interbank networks \sep Air-traffic networks
\end{keywords}

\maketitle
\setlength{\parindent}{0pt}
\setlength{\parskip}{2pt plus 0.5pt minus 0.5pt}

\section{Introduction}
Networks provide a natural way to capture the interactions among the units of complex real-world systems. Yet, a central insight of the recent literature is that these systems are often shaped not only by pairwise ties, but also by higher-order interactions, see, e.g., \cite{Milo2002,benson2018simplicial, battiston2021physics}. For example, latent mechanisms such as homophily (the tendency of similar nodes to be connected) or transitivity (the tendency of two nodes to be connected if they share a common neighbor) may involve not only pairs but groups of nodes, explaining an \emph{overexpression} of higher-order link structures, such as triangles, wedges, and more general motifs, with respect to standard benchmarks, see, e.g., \cite{peixoto2022disentangling,sarker2024higher}. This recognition has shifted attention from dyadic structure to group-level dependencies, motivating a growing literature on higher-order networks, hypergraphs, and simplicial complexes, and highlighting their relevance for both structural description and dynamical phenomena. Using the lens of higher-order networks, many systems---financial, social, technological, and transportation---have been recognized to display robust meso- and higher-order organization, including triadic closure, motifs, and clustering, that cannot generally be reproduced by standard models based on dyadic statistics, see, e.g.,  \cite{Newman2010,Barabasi2016,Rapoport1953,Milo2002}. This limitation poses a central challenge in probabilistic network modeling, which underpins the standard null models used in data analysis. In practice, statistical inference is typically carried out using models that are both \emph{analytically tractable} and \emph{parsimonious} in their parameterization, which often leads to factorized probability distributions based on linear network statistics such as the number of links or node degrees, see, e.g, \cite{Jaynes1957,ParkNewman2004,BoydVandenberghe2004,GarlaschelliLoffredo2008,WainwrightJordan2008,ChatterjeeDiaconisSly2011}. This choice creates a common failure mode: a model can match its \emph{intended} linear constraints (e.g., edge density, degrees, block densities) while remaining systematically inconsistent with simple structural diagnostics such as wedge and triangle counts. However, as soon as nonlinear network statistics are included, the model's likelihood does {\it not} admit a closed-form expression anymore and analytical tractability is lost.
The present work proposes a pragmatic remedy at the \emph{inference} level.
Rather than changing the generative model by introducing explicit triadic or higher-order terms---which typically induce complex dependence structures, prevent analytical tractability, destroy convexity, and make the estimation infeasible, see, e.g., \cite{FrankStrauss1986,WassermanPattison1996,SnijdersEtAl2006,Handcock2003,Schweinberger2011}---we keep a tractable dyad-independent likelihood and \emph{penalize} it by standardized discrepancies in targeted motifs.
This yields \emph{Penalized Likelihood with Structural Discrepancies} (PLSD): a likelihood objective augmented by a dimensionless motif penalty that can be tuned (when replicate graphs are available) or calibrated (when only a single network snapshot is observed).

From an empirical perspective, core-periphery networks are an especially natural setting for studying the overexpression of higher-order structures, because cores are typically the part of the network where density, clustering, and triadic closure are most pronounced, whereas peripheries are comparatively sparse and structurally simpler. This contrast makes core-periphery systems a concrete and scientifically relevant testbed for assessing how higher-order patterns emerge and how they affect network inference, a point that is well illustrated by the emergence results of \cite{verma2016emergence} and by research work on core-periphery detection and modeling, see, e.g., \cite{BorgattiEverett2000,RombachPorterFowlerMucha2017,ShenHanLiWongPeng2021}. Particular examples are interbank markets: persistent cores and tiering have been documented and linked to systemic fragility, see, e.g., \cite{squartini2013early,FrickeLux2015,CraigVonPeter2014,BaruccaLillo2016,BaruccaLillo2018,SoramakiBechArnoldGlassBeyeler2007,XuHeLi2016}; transportation networks also display hub concentration and clustering, central characteristics for performance and robustness, see, e.g., \cite{BarratBarthelemyPastorSatorrasVespignani2004,GuimeraMossaTurtschiAmaral2005,Bagler2008,Newman2010}.
The inferred organization of an interbank network is itself sensitive to the benchmark and observation scale. In particular, Barucca and Lillo \cite{BaruccaLillo2016} show for eMID that degree heterogeneity, the choice between the stochastic block model and its degree-corrected counterpart, and temporal aggregation can change whether a bipartite or core--periphery structure is selected. Their analysis concerns model selection for the large-scale organization of the network. PLSD addresses a complementary question: conditional on a tractable dyad-independent model, it corrects structural misspecification by directly targeting wedge and triangle discrepancies and quantifies the resulting likelihood cost.
These domains also illustrate an important point: the \emph{same} structural penalty can have different predictive costs depending on the data-generating mechanisms and scale, so empirical evaluation must quantify the trade-off rather than assume it away.

\paragraph*{Contributions.}
\begin{itemize}
\item \textbf{PLSD methodology.} We formalize PLSD as a standardized discrepancy penalty for wedges and triangles that corrects structural misspecification without changing the selected dyad-independent model family, introduce a dimensionless scaling that makes $\lambda$ comparable across datasets, and derive a small-$\lambda$ linear response formula that makes the direction of the correction explicit.
\item \textbf{Controlled simulations.} Using a toy core--periphery model, we show that PLSD is inert under correct specification and becomes active under two targeted misspecifications (triadic closure in the core and heterogeneous core activity), yielding large motif gains at small held-out likelihood cost.
\item \textbf{Core--periphery application.} We develop a mixed-constraint maximum-entropy core--periphery ERGM used for empirical inference, and complement it with a Pareto-marked unconditional layer that yields semi-closed motif formulas and a clear sparse-regime scaling interpretation via a big-jump mechanism \cite{Resnick2007,VezzaniBarkaiBurioni2019}.
\item \textbf{Empirical evidence.} On eMID and US air traffic temporal networks, we use the core--periphery ERGM introduced in Sec.~\ref{sec:cp_model}, estimated both by plain Negative Log-Likelihood (NLL) and by PLSD so that the effect of the penalty can be isolated. PLSD markedly improves wedge and triangle agreement and quantifies the link-level trade-off in two domains and scales.
\end{itemize}

The remainder of the paper is organized as follows. Section \ref{sec:plsd} introduces the PLSD inference methodology. Section \ref{sec:toy} validates the proposed PLSD approach through Monte Carlo simulations of a toy stochastic block model. Section \ref{sec:cp_model} presents a new core-periphery ERGM, while Section \ref{sec:pareto} derives analytical results on the scaling properties of network metrics. Section \ref{sec:empirical} reports two empirical applications to interbank markets and air traffic networks, providing further evidence in support of the proposed methodology. Finally, Section \ref{sec:conclusions} concludes the paper and discusses possible extensions, while the Appendix contains complementary derivations and additional results.

\clearpage

\section{Penalized likelihood with structural discrepancies (PLSD)}
\label{sec:plsd}

\subsection{Motif targets and standardized discrepancies}
Let $A$ be the adjacency matrix of an undirected, unweighted simple graph with $N$ vertices.
We focus on three basic subgraph counts:
the number of edges $L$, the number of length-2 paths (wedges) $W$, and the number of triangles $T$.
Following standard motif practice \cite{Newman2010,Milo2002}, we define:
\begin{equation}
\label{eq:motifs}
\begin{aligned}
L &= \sum_{i<j} A_{ij} \\
W &= \sum_i \sum_{\substack{j<k\\ j,k\neq i}} A_{ij}A_{ik} \\
T &= \sum_{i<j<k} A_{ij}A_{ik}A_{jk}
\end{aligned}
\end{equation}

Consider a dyad-independent model with probability matrix $P=(p_{ij})$ and parameters $\theta$ (possibly including node-level fields).
For each motif $M\in\{W,T\}$, denote by $\mu_M(\theta)=\E_\theta[M]$ its model expectation and by
$\sigma_M(\theta)=\sqrt{\Var_\theta(M)}$ its standard deviation under the model.
The \emph{standardized discrepancy} (motif Z-score) is:
\begin{equation}
Z_M(\theta) = \frac{M_{\mathrm{obs}} - \mu_M(\theta)}{\sigma_M(\theta)}, 
\qquad 
M\in\{W,T\}
\end{equation}
where $M_{\mathrm{obs}}$ is the observed motif count.
In the dyad-independent case, $\mu_M$ and $\Var(M)$ can be computed analytically by combinatorial sums of products of $p_{ij}$ (and stabilized with small ridges when the variance is tiny).
In empirical fits, we evaluate $\sigma_M(\theta)$ at the maximum-likelihood estimator (MLE) and \emph{freeze} it during PLSD optimization to reduce computational cost and avoid instability when $\Var(M)$ varies rapidly with $\theta$.

\subsection{Diagonal PLSD objective and scale normalization}
Let $\Phi(\theta)$ be the negative log-likelihood (NLL) of the model.
The diagonal PLSD estimator minimizes:
\begin{equation}
\label{eq:PLSD}
\Phi_{\lambda}(\theta) 
= 
\Phi(\theta) 
+ 
\lambda\,\frac{S_{\Phi}}{S_Z}\left(Z_W(\theta)^2 + Z_T(\theta)^2\right)
\end{equation}
where $\lambda\ge 0$ is a scalar penalty strength and the scale factors:
\begin{equation}
S_{\Phi}=\Phi(\hat\theta_0), 
\qquad 
S_Z=Z_W(\hat\theta_0)^2+Z_T(\hat\theta_0)^2
\end{equation}
are evaluated at the unpenalized MLE $\hat\theta_0$.
When $S_Z$ is close to zero (which occurs under correct specification in simulations), we use the stabilized choice $S_Z\leftarrow\max(S_Z,\varepsilon)$ with $\varepsilon>0$.
The ratio $S_\Phi/S_Z$ makes $\lambda$ dimensionless and improves comparability across datasets and time snapshots.

The penalty in Eq.~\eqref{eq:PLSD} can be read as a \emph{minimum-distance} correction to the likelihood surface: $Z_W^2+Z_T^2$ measures a standardized moment mismatch, akin to the diagonal part of a generalized method-of-moments criterion \cite{Hansen1982}.
Unlike pure moment-based fitting, however, PLSD preserves the likelihood as the primary objective and uses $\lambda$ to control the extent of the distortion.

\subsection{Small-$\lambda$ linear response}
Because $\hat\theta_0$ solves $\nabla\Phi(\hat\theta_0)=0$, the penalized estimate admits a transparent first-order expansion for small $\lambda$.
Write $H(\hat\theta_0)=\nabla^2\Phi(\hat\theta_0)$ and
$g_{\mathrm{pen}}(\theta)=\nabla\bigl(Z_W(\theta)^2+Z_T(\theta)^2\bigr)$.
A first-order Taylor expansion of the stationarity condition
$\nabla\Phi_{\lambda}(\hat\theta_{\lambda})=0$ yields:
\begin{equation}
\Delta\theta \equiv \hat\theta_{\lambda}-\hat\theta_0 
\approx 
-\lambda\,\frac{S_{\Phi}}{S_Z}\,H(\hat\theta_0)^{-1}\,g_{\mathrm{pen}}(\hat\theta_0)
\label{eq:linear_response}
\end{equation}
so that PLSD moves the estimate along directions where (i) the likelihood surface is flat (large $H^{-1}$) and (ii) the standardized motif mismatch is most sensitive.
Appendix~\ref{sec:appendix_linear} reports a short derivation and verifies Eq.~\eqref{eq:linear_response} against Monte Carlo estimates in our toy simulations.

\subsection{Choosing $\lambda$: tuning in simulations vs calibration in empirical snapshots}
\paragraph*{Simulation regime (replicate graphs available).}
When the data-generating process can be sampled repeatedly, $\lambda$ is a genuine tuning parameter.
For each replicate, we fit PLSD on an independent \emph{training} graph and evaluate on an independent \emph{test} graph:
(i) the held-out NLL, $\NLL^{\mathrm{test}}(\lambda)$, and
(ii) the held-out motif mismatch
$Z_{\mathrm{pen}}^{\mathrm{test}}(\lambda)=Z_W^{\mathrm{test}}{}^2+Z_T^{\mathrm{test}}{}^2$.
To make the trade-off comparable across replicates, we normalize by the $\lambda=0$ values on the same test graph:
\[
\NLLnorm(\lambda)=\frac{\NLL^{\mathrm{test}}(\lambda)}{\NLL^{\mathrm{test}}(0)},
\qquad
\Zpennorm(\lambda)=\frac{Z_{\mathrm{pen}}^{\mathrm{test}}(\lambda)}{Z_{\mathrm{pen}}^{\mathrm{test}}(0)}
\]
and select $\lambda$ by minimizing the balanced score:
\[
J(\lambda)=w\,\NLLnorm(\lambda)+(1-w)\,\Zpennorm(\lambda),
\qquad
w=\tfrac12
\]
This protocol is meaningful precisely because independent test graphs exist.

\paragraph*{Empirical regime (one network per snapshot).}
For empirical temporal networks, each snapshot provides a single observed graph.
Within-snapshot train/test splitting would discard data and introduce unnecessary variance, while time-based splits would require strong stationarity assumptions and can be unstable when the number of snapshots per regime is limited.
Accordingly, in the empirical analyses we do not tune $\lambda$ for prediction.
Instead, we \emph{calibrate the scale} of the discrepancy term so that it is commensurate with the likelihood within each snapshot.

Concretely, within each snapshot we first scan candidate core sizes $m\in\mathcal{M}$ (Sec.~\ref{sec:empirical}) and, for each $m$, compute:
the MLE NLL $\Phi_m$ and the squared standardized residuals $Z_{T,m}^2$ and $Z_{W,m}^2$.
We then set motif-specific penalty weights:
\[
\lambda_T = \frac{\overline{\Phi}}{\widehat{\sigma}_T},
\qquad
\lambda_W = \frac{\overline{\Phi}}{\widehat{\sigma}_W}
\]
where $\overline{\Phi}$ is the mean of $\{\Phi_m\}_{m\in\mathcal{M}}$ and $\widehat{\sigma}_M$ is a robust scale estimate (the median absolute deviation, MAD, scaled by $1.4826$, with standard-deviation fallback if MAD is zero) of $\{Z_{M,m}^2\}_{m\in\mathcal{M}}$ \cite{RousseeuwCroux1993}.
With this choice, a one-scale change of each discrepancy across plausible core sizes contributes on the order of $\overline{\Phi}$ to the objective, preventing the penalty from becoming dominant (or negligible) due to units alone.

For reporting, we map $(\lambda_T,\lambda_W)$ to an \emph{effective} dimensionless scalar $\lambda_{\mathrm{eff}}$ associated with Eq.~\eqref{eq:PLSD}.
Define $\lambda_{\mathrm{eff}}$ as the unique scalar satisfying, at the MLE:
\[
\lambda_{\mathrm{eff}}\frac{S_\Phi}{S_Z}\bigl(Z_T^2+Z_W^2\bigr)
=
\lambda_T Z_T^2+\lambda_W Z_W^2
\]
Solving gives:
\[
\lambda_{\mathrm{eff}}
=
\frac{S_Z}{S_\Phi}\,
\frac{\lambda_T Z_T^2+\lambda_W Z_W^2}{Z_T^2+Z_W^2}
\]
This scalar is used only as a normalized summary to compare snapshots/datasets; inference itself uses the calibrated $(\lambda_T,\lambda_W)$.

\subsection{Optimization}
Even when $\Phi(\theta)$ is convex, adding $Z_M(\theta)^2$ generally breaks convexity, so PLSD must be handled with robust numerical optimization.
Throughout, we use bound-constrained quasi-Newton optimization (L-BFGS-B) initialized at the unpenalized MLE and continued in $\lambda$ when scanning a grid \cite{ByrdLuNocedalZhu1995,NocedalWright2006}.
This warm-start continuation is crucial in practice because the penalty can introduce additional local minima.

\section{Simulations: toy core--periphery model with triadic closure}
\label{sec:toy}

We now want to isolate the role of PLSD in a fully controlled setting.
We generate synthetic graphs from (i) a well-specified dyad-independent core--periphery model and (ii) a targeted misspecification that introduces triadic closure inside the core.
Because independent replicate graphs are available, $\lambda$ can be tuned by the train/test protocol of Sec.~\ref{sec:plsd}.

\subsection{Toy model and baseline specification}
We consider a flat core--periphery network on $N=2000$ nodes with a fixed core of size $N_c=300$ and periphery size $N_p=1700$.
Edges are independent with probabilities:
\begin{equation}
 p_{ij} = \sigmoid\left(y + x\bigl(\I[i\in C] + \I[j\in C]\bigr)\right)
\end{equation}
so that $p_{cc}=\sigmoid(y+2x)$, $p_{cp}=\sigmoid(y+x)$, and $p_{pp}=\sigmoid(y)$.
The baseline (well-specified) case uses $(x^*,y^*)=(1.2,-5.0)$ for both data generation and fitting.
We enforce $x\ge 0$ so that the ordering $p_{pp}\le p_{cp}\le p_{cc}$ is preserved.

\subsection{Targeted misspecification: triadic closure in the core}
To create a minimal and interpretable departure from dyad-independence, we inject explicit triadic closure \emph{only} among core nodes.
Starting from a baseline draw, we repeat $\texttt{triad\_steps}=100000$ times:
sample a triplet of distinct core nodes $(i,j,k)$ uniformly at random; if exactly two of the three core--core edges are present, add the missing edge; otherwise do nothing.
This procedure preserves symmetry, avoids self-loops, and increases clustering primarily by closing length-2 paths into triangles.
Table~\ref{tab:triadic_stats} reports the induced densification in the core--core block and the mean number of edges added.
Specifically, $p_{cc,0}$ is the baseline core--core edge probability under the dyad-independent model, $p_{cc,\mathrm{marg}}$ is the marginal core--core density \emph{after} closure (mean and s.d.\ over $n_{\mathrm{samples}}$ trials), and ``edges added'' reports the average number of closed edges added by the procedure; $\texttt{triad\_steps}$ and $n_{\mathrm{samples}}$ are the fixed Monte Carlo settings.

\begin{table}[pos=H]
\centering
\caption{Triadic closure calibration within the core block (Monte Carlo).}
\label{tab:triadic_stats}
\begin{tabular}{lc}
\toprule
Quantity & Value \\
\midrule
triad\_steps & 100000 \\
n\_samples & 100 \\
$p_{cc,0}$ & 0.0691 \\
$p_{cc,\mathrm{marg}}$ & 0.1188 (0.0041) \\
edges added & 2229.35 (135.42) \\
\bottomrule
\end{tabular}
\end{table}

\subsection{PLSD fitting, selection, and results}
\begin{figure}[pos=H]
\centering
\begin{subfigure}{0.49\columnwidth}
\centering
\includegraphics[width=\linewidth]{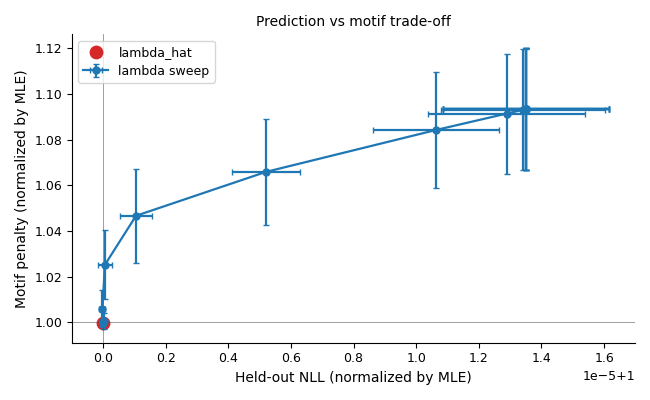}
\caption{Well-specified baseline: no material trade-off near $\lambda=0$.}
\end{subfigure}
\begin{subfigure}{0.49\columnwidth}
\centering
\includegraphics[width=\linewidth]{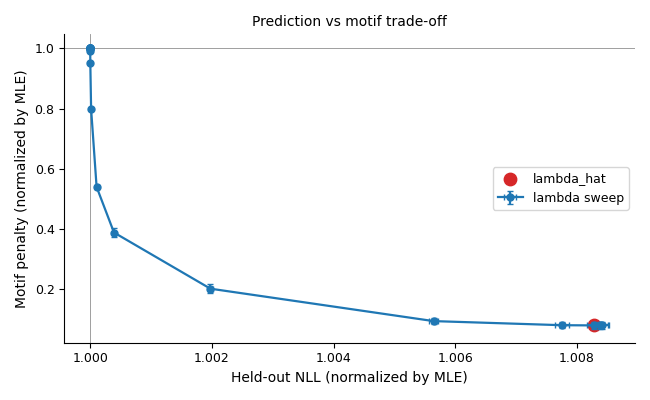}
\caption{Triadic closure misspecification: strong motif improvement at small NLL cost.}
\end{subfigure}
\caption{Toy-model trade-off curves for the well-specified and triadic-closure scenarios.}
\label{fig:toy_tradeoff}
\end{figure}

\begin{figure}[pos=H]
\centering
\begin{subfigure}{0.95\columnwidth}
\centering
\includegraphics[width=\linewidth]{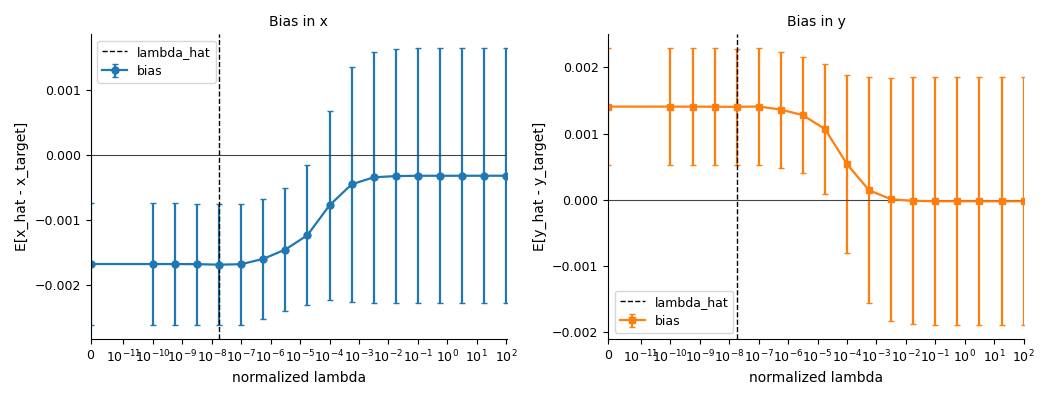}
\caption{Baseline: parameter bias vs $\lambda$, measured relative to the data-generating parameter. Small apparent improvements away from $\lambda=0$ are Monte Carlo fluctuations in held-out evaluation rather than systematic bias correction.}
\end{subfigure}
\vspace{0.4em}
\begin{subfigure}{0.95\columnwidth}
\centering
\includegraphics[width=\linewidth]{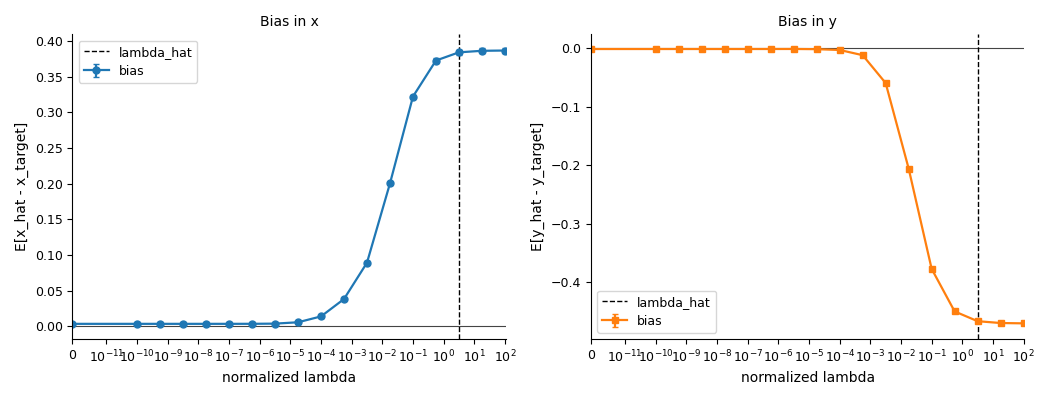}
\caption{Triadic closure: parameter bias vs $\lambda$.}
\end{subfigure}
\caption{Parameter bias under PLSD in the toy simulation.}
\label{fig:toy_bias}
\end{figure}

We run $R=100$ Monte Carlo replicates.
In each replicate, we draw one training graph and $K=10$ independent test graphs from the same underlying parameters; in the misspecified scenario, the closure procedure is applied independently to each draw.
For each $\lambda$ on a predefined grid, we fit PLSD on the training graph (L-BFGS-B with $x\ge 0$), evaluate $\NLL$ and the motif penalty on the test graph, and select $\hat\lambda$ by $J(\lambda)$ with $w=1/2$.
We report normalized test NLL, normalized Z-penalty, the combined score $J(\lambda)$, and parameter bias in Table~\ref{tab:simulation_baseline_vs_triadic} and Figs.~\ref{fig:toy_tradeoff}--\ref{fig:toy_bias}.

\begin{table}[pos=H]
\caption{Balanced $\lambda$ selection summary for the toy model. Normalized metrics and $J(\hat\lambda)$ are evaluated on held-out graphs; values are mean (SE) over replicates. $\text{NLL}_{\text{norm}}$ and $\text{Zpen}_{\text{norm}}$ are normalized by their $\lambda=0$ test values.}
\label{tab:simulation_baseline_vs_triadic}
\small
\begin{tabular}{lcc}
\toprule
Metric & Baseline & Misspecified \\
\midrule
$\hat\lambda$ (norm.) & 1.78e-08 & 3.16 \\
$\text{NLL}_{\text{norm}}(\hat\lambda)$ & 1.000 (0.000) & 1.008 (0.000) \\
$\text{Zpen}_{\text{norm}}(\hat\lambda)$ & 1.000 (0.001) & 0.081 (0.011) \\
$J(\hat\lambda)$ & 1.000 (0.001) & 0.544 (0.005) \\
$\hat x$ & 1.198 (0.001) & 1.829 (0.003) \\
$\hat y$ & -4.999 (0.001) & -5.563 (0.003) \\
\bottomrule
\end{tabular}
\end{table}

\FloatBarrier

Two qualitative behaviors emerge.
First, under correct specification, $\hat\lambda$ concentrates near $0$, and the penalty term is effectively inactive: the MLE already reproduces wedge and triangle counts up to sampling noise.
Second, under triadic closure, PLSD selects a \emph{strictly positive} penalty.
Quantitatively, the selected $\hat\lambda$ raises $\NLLnorm$ to $1.008$ while reducing $\Zpennorm$ to $0.081$ (Table~\ref{tab:simulation_baseline_vs_triadic}), i.e., a sub-$1\%$ predictive cost for roughly a $92\%$ reduction in motif mismatch.
Additional diagnostics, including variance/MSE curves and Z-score trajectories as functions of $\lambda$, are reported in Appendix~\ref{sec:appendix_diagnostics}.
A second misspecification case---heterogeneous core fitness fitted by a flat-core model---is reported in Appendix~\ref{sec:appendix_hetero}; it shows the same qualitative PLSD response.
In the well-specified baseline bias panel, $x_{\mathrm{target}}$ and $y_{\mathrm{target}}$ are the data-generating parameters.
Thus $\lambda=0$ is the theoretical reference point for unbiased likelihood estimation; the small apparent dips at nonzero $\lambda$ in Fig.~\ref{fig:toy_bias} are within Monte Carlo standard errors and are not interpreted as systematic bias correction.
\FloatBarrier

\subsection{Recovery of the core size}
\label{sec:toy_core_size}
\begin{figure}[pos=H]
\centering
\begin{subfigure}{0.49\columnwidth}
\centering
\includegraphics[width=\linewidth]{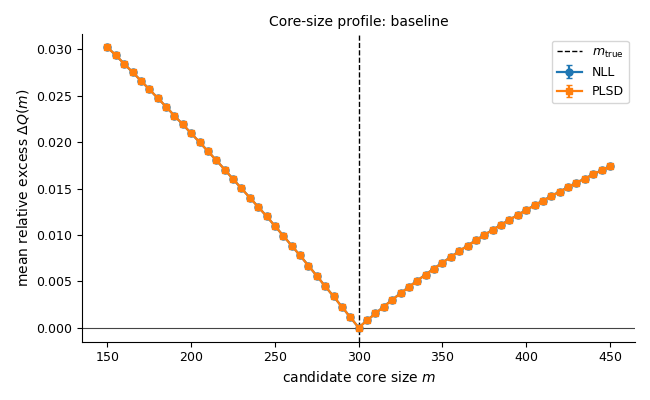}
\caption{Well-specified baseline.}
\end{subfigure}
\begin{subfigure}{0.49\columnwidth}
\centering
\includegraphics[width=\linewidth]{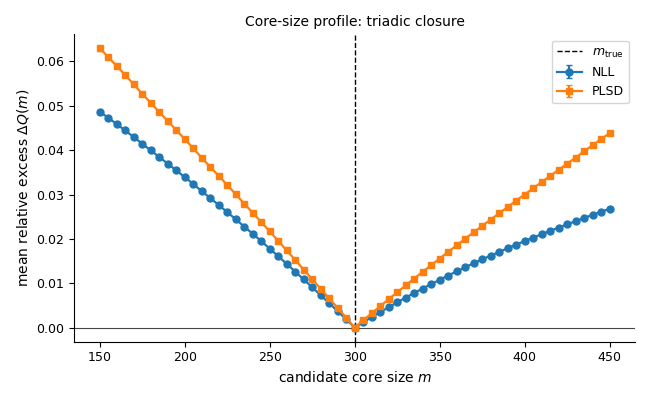}
\caption{Triadic closure misspecification.}
\end{subfigure}
\caption{Core-size criterion profiles at $x^*=1.2$. For each replicate and
criterion $Q$, we report
$\Delta Q(m)=[Q(m)-\min_{m'}Q(m')]/|\min_{m'}Q(m')|$, averaged over 100
replicates with 95\% Monte Carlo intervals. The vertical line marks
$m_{\mathrm{true}}=300$. The PLSD curve is the penalized selection objective,
not a likelihood. Both NLL and PLSD attain their minimum at the true core size
in all replicates, under both baseline generation and triadic closure. The figure is retained as a diagnostic of minimum location; the recovery experiment in Fig.~\ref{fig:toy_core_size_recovery} provides the main performance comparison over weak-to-strong separation regimes.}
\label{fig:toy_core_size_profiles}
\end{figure}

\noindent
The preceding experiments treat the true core membership and size as known.
We now test whether the core size itself can be recovered by NLL and PLSD.
To isolate size identification from node-ranking errors, candidate cores are
nested in oracle order: true core nodes appear first, followed by periphery
nodes. For each candidate size $m$, the flat toy model is refitted and the
selected size minimizes either the NLL or the PLSD objective. Since both fits
contain the same two parameters $(x,y)$ for every $m$, no AIC correction is
needed in this experiment.
\\
For PLSD, a normalized $\lambda$ is selected at $m_{\mathrm{true}}=300$ using
50 independent calibration replicates and the train/test protocol above, with
10 test graphs per replicate. The selected normalized $\lambda$ and its
graph-specific scale factor are then held fixed throughout each scan over
$m$. Evaluation uses a separate set of 100 replicates. This is therefore a
controlled recovery experiment rather than a fully data-driven procedure for
unknown empirical cores.
\\
We first inspect the complete selection profiles at the reference value
$x^*=1.2$. For each candidate size, both criteria are expressed as relative
excess above their own minimum, so the location and sharpness of the minima
can be compared despite the different scales of the NLL and PLSD objectives.
If PLSD preserves the likelihood-based core detection, its profile may be
distorted by the motif penalty under misspecification, but its minimum should
remain aligned with $m_{\mathrm{true}}$.
Fig.~\ref{fig:toy_core_size_profiles} reports these criterion profiles at the
reference value $x^*=1.2$.

\FloatBarrier

\begin{figure}
\centering
\begin{subfigure}{0.49\columnwidth}
\centering
\includegraphics[width=\linewidth]{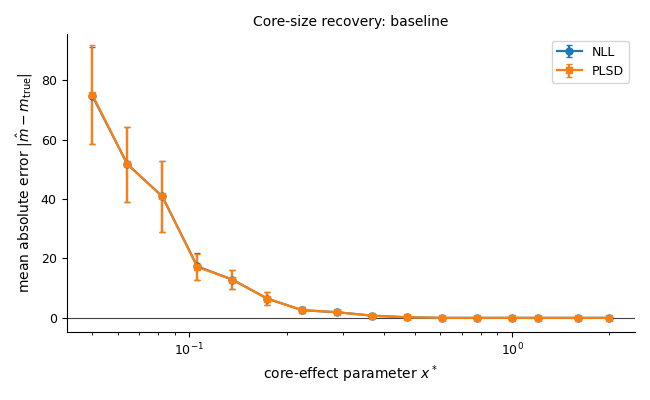}
\caption{Well-specified baseline.}
\end{subfigure}
\vspace{0.4em}
\begin{subfigure}{0.49\columnwidth}
\centering
\includegraphics[width=\linewidth]{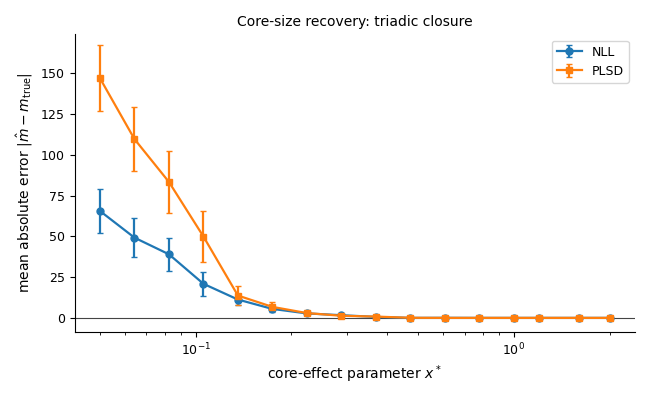}
\caption{Triadic closure misspecification.}
\end{subfigure}
\caption{Mean absolute core-size error as structural separation increases.
The horizontal axis uses a logarithmic grid below $x^*=1$ to resolve weak
core--periphery separation, followed by reference values at $x^*=1.2,1.6,2.0$.
Candidate sizes are searched coarse-to-fine and error bars are 95\% Monte
Carlo intervals for the mean over 100 independent evaluation graphs. Recovery
is most difficult in the weakest-separation regime and becomes essentially
exact once the core is clearly identifiable. Under triadic closure, PLSD and
NLL remain aligned in the strong-signal regime, while the penalized criterion
can be less stable than NLL when separation is extremely weak and the
core-size profile is nearly flat.}
\label{fig:toy_core_size_recovery}
\end{figure}

\noindent
We next vary $x^*$ from $0.2$ to $2.0$ while keeping $y^*=-5$. Although we
refer to $x^*$ as the core-density parameter, it acts on the logit scale:
$p_{cc}=\sigmoid(y^*+2x^*)$. Hence increasing $x^*$ increases the separation
between core and periphery and should improve identification of the change
point at $m_{\mathrm{true}}$.
Fig.~\ref{fig:toy_core_size_recovery} summarizes this recovery experiment over
weak-to-strong separation regimes.

\FloatBarrier

\noindent
These results show that PLSD can be used as a \emph{core-size selection
criterion} without systematically displacing the change point identified by
the likelihood once the core--periphery signal is identifiable. In the
well-specified baseline this agreement is expected in the moderate- and
strong-signal regimes because the selected penalty is essentially inactive.
Under very weak separation, however, the profile is nearly flat and even a
small structural penalty can amplify finite-sample variability in the selected
size. Under triadic closure, the motif penalty materially changes the shape
and scale of the profile, but its minimum remains at $m_{\mathrm{true}}$ once
the signal is not dominated by noise. Thus the structural correction does not
systematically purchase motif agreement by expanding or shrinking the core in
this controlled setting.
\\
The weak-separation points are descriptive rather than evidence of a general
superiority of either criterion: confidence intervals are large and both
methods become essentially exact once core--periphery separation is moderate.
More importantly, this experiment identifies only the size of a core along a prescribed nested
sequence. It does not test recovery of the node ranking, arbitrary changes in
membership, or joint combinatorial optimization over all possible cores.
Accordingly, it supports the empirical pipeline in which a ranking is
constructed first and PLSD replaces NLL only in the subsequent scan over
$m$. Ranking uncertainty and the AIC/BIC complexity correction remain
separate components of the core-detection procedure.

\FloatBarrier

\section{A core--periphery ERGM}
\label{sec:cp_model}

We now turn to the core--periphery model used in empirical inference and in the theoretical calculations that follow.
The construction is based on the maximum entropy principle: we choose constraints that are (i) directly interpretable in a core--periphery setting and (ii) yield dyad-independence and convex likelihood inference.
The unconditional heavy-tailed (Pareto) layer is introduced separately in Sec.~\ref{sec:pareto} as a theoretical device.

\subsection{Conditional inference layer: maximum-entropy with mixed constraints}
Consider an undirected simple graph $A$ on vertex set $V$, and let $C\subset V$ denote a candidate core (with periphery $P=V\setminus C$).
We impose the mixed constraints:
\begin{equation}
\begin{aligned}
\E\left[\sum_{i<j} A_{ij}\right] &= L^*\\
\E\left[\sum_{j\neq i} A_{ij}\right] &= k_i^*, \quad i\in C
\end{aligned}
\end{equation}
i.e., we match global density and the degrees of core nodes, while leaving periphery degrees unconstrained.
By the maximum entropy principle \cite{Jaynes1957,GarlaschelliLoffredo2008}, the resulting ERGM is dyad-independent with logistic edge probabilities:
\begin{equation}
\mathrm{logit}(p_{ij}) = y + x_i\I[i\in C] + x_j\I[j\in C]
\end{equation}
where $y$ is a global field and $(x_i)_{i\in C}$ are core-specific fields (set to $0$ in the periphery).
This is a core-restricted version of the ``soft'' configuration-type ensembles widely used for network reconstruction and null modeling \cite{ParkNewman2004,GarlaschelliLoffredo2008,SquartiniGarlaschelli2011,ChatterjeeDiaconisSly2011}.

The negative log-likelihood is:
\begin{equation}
\begin{aligned}
\Phi(x,y)&=\sum_{i<j}\left[\log\left(1+e^{\lambda_{ij}}\right)-A_{ij}\lambda_{ij}\right]\\
\lambda_{ij}&=y+x_i\I[i\in C]+x_j\I[j\in C]
\end{aligned}
\end{equation}
and the gradients enforce moment matching:
\begin{equation}
\begin{aligned}
\frac{\partial \Phi}{\partial y} &= \E_\theta[L]-L_{\mathrm{obs}} \\
\frac{\partial \Phi}{\partial x_i} &= \E_\theta[k_i]-k_{i,\mathrm{obs}}, \quad i\in C
\end{aligned}
\end{equation}

\subsection{Convexity, identifiability, and constraints}
Because $\Phi$ is a sum of log-partition terms minus a linear term, it is convex in $(y,x_i)$ \cite{BoydVandenberghe2004,WainwrightJordan2008}.
Strict convexity holds on the interior when $0<p_{ij}<1$ and the periphery is nonempty, implying a unique MLE.
To preserve a core--periphery ordering, we impose $x_i\ge 0$ for all $i\in C$, which guarantees $p_{pp}\le p_{cp}\le p_{cc}$ in the flat-core limit and prevents ``anti-core'' solutions.
The resulting bound-constrained convex program still has a unique solution (allowing boundary optima with some $x_i=0$).
When PLSD is applied, convexity is lost and we revert to robust bound-constrained methods with warm starts along the penalty grid, as in Sec.~\ref{sec:plsd}.

\subsection{Motif moments under dyad independence}
For a fixed probability matrix $P=(p_{ij})$, motifs are defined as in Eq.~\eqref{eq:motifs}, and the conditional expectations are:
\begin{equation}
\begin{aligned}
\E[L\mid P] &= \sum_{i<j} p_{ij} \\
\E[W\mid P] &= \sum_i \sum_{\substack{j<k\\ j,k\neq i}} p_{ij}p_{ik}\\
\E[T\mid P] &= \sum_{i<j<k} p_{ij}p_{ik}p_{jk}
\end{aligned}
\end{equation}
For PLSD, we also require variance proxies $\Var(W)$ and $\Var(T)$ under the dyad-independent ensemble.
In simulations, these are computed analytically and can be evaluated exactly.
In empirical fits, the analytic expressions remain available but can be numerically delicate when motif variability is tiny; we therefore stabilize the denominators with small ridges and, unless otherwise stated, freeze $\sigma_W$ and $\sigma_T$ at their MLE values during penalized optimization.

\section{Unconditional Pareto layer and asymptotic regimes}
\label{sec:pareto}

This section introduces an \emph{unconditional} layer that is used only for analytical scaling results.
The inference model of Sec.~\ref{sec:cp_model} treats the core fields $(x_i)$ as fixed parameters.
Here, by contrast, we model them as i.i.d.\ heavy-tailed marks and integrate them out.
This produces non-independent edges, but it yields tractable low-dimensional integral formulas and clarifies the origin of sparse-regime scaling in motifs and degrees.

\subsection{Pareto marks}
We endow core nodes with i.i.d.\ Pareto marks $X$ with density:
\begin{equation}
p(x)=C_\alpha x^{-\alpha}\I[x\ge x_{\min}], 
\qquad 
C_\alpha=(\alpha-1)x_{\min}^{\alpha-1}
\end{equation}
with $\alpha>1$ to ensure a finite mean.
Conditionally on marks $X_i,X_j$, a core--core edge occurs with probability $\sigmoid(y+X_i+X_j)$, while a core--periphery edge uses only one mark, $\sigmoid(y+X_i)$, and periphery--periphery edges have probability $\sigmoid(y)$.
After marginalization, edges are no longer independent, but block probabilities and motif kernels can be written as low-dimensional integrals.

\subsection{Semi-closed block probability formulae}
We display explicitly the core--core block probability, since it is the template for all other cases.
Conditionally on $X_1,X_2$, a core--core edge occurs with probability $\sigmoid(y+X_1+X_2)$, hence:
\begin{equation}
\begin{aligned}
&
P_{CC}(y) = C_\alpha^2 \cdot \\
&
\cdot
\int_{x_{\min}}^{\infty}\int_{x_{\min}}^{\infty} x_1^{-\alpha}x_2^{-\alpha}\sigmoid(y+x_1+x_2)\,dx_1\,dx_2
\end{aligned}
\end{equation}
To reduce the two-dimensional integral, we use the Laplace representation:
\begin{equation}
\sigmoid(u)=\int_0^{\infty}e^{-q}e^{-q e^{-u}}\,dq
\end{equation}
apply Fubini--Tonelli, and change variables to
$s=x_1+x_2$ and $t=x_1-x_2$.
The domain becomes $s\ge 2x_{\min}$ with $|t|\le s-2x_{\min}$, and the inner integral reduces to an incomplete Beta term, yielding:
\begin{equation}
\begin{aligned}
&
P_{CC}(y) = 2^{2\alpha-1}C_\alpha^2 \cdot \\
& \cdot
\int_{2x_{\min}}^{\infty}\sigmoid(s+y)\,s^{1-2\alpha}
B_{\lambda(s)^2}\!\left(\tfrac{1}{2},1-\alpha\right)\,ds
\end{aligned}
\end{equation}
where $\lambda(s)=1-2x_{\min}/s$ and $B_x(a,b)=\int_0^x t^{a-1}(1-t)^{b-1}\,dt$.
Finally, the affine change $z=\lambda(s)$ maps the domain to $(0,1)$ and yields the compact one-dimensional form:
\begin{equation}
\begin{aligned}
&
P_{CC}(y)=2(\alpha-1)^2 \cdot \\
& \cdot
\int_{0}^{1}(1-z)^{2\alpha-3}
B_{z^2}\!\left(\tfrac{1}{2},1-\alpha\right)\sigmoid\!\left(\frac{2x_{\min}}{1-z}+y\right)\,dz
\label{eq:pcc_z}
\end{aligned}
\end{equation}
The remaining block probabilities and motif kernels follow by analogous one- or two-dimensional integrals; explicit formulas are collected in Appendix~\ref{sec:appendix_unconditional}.

\clearpage
\subsection{Motif regimes and sparse-regime scaling}

\paragraph*{Dense regime.}
As $y\to+\infty$, $\sigmoid(u)\to 1$ uniformly, so all block probabilities and motif kernels converge to $1$.

\paragraph*{Linear-response regime.}
When $y$ is tuned so that the typical activation variables remain $O(1)$ for the bulk of pairs:
\[
z \equiv x+x'+y = O(1)
\]
the logistic can be expanded around $z=0$.
Writing $\sigmoid(z)=\bigl(1+e^{-z}\bigr)^{-1}$ and using $e^{-z}=1-z+\tfrac12 z^2+O(z^3)$ gives:
\[
\sigmoid(z)
=
\frac{1}{2-z+\tfrac12 z^2+O(z^3)}
=
\frac12+\frac14 z+O(z^3)
\]
i.e.:
\begin{equation}
\sigmoid(z)=\tfrac12+\tfrac14 z+O(z^3)
\label{eq:sigmoid_linear}
\end{equation}
Inserting Eq.~\eqref{eq:sigmoid_linear} into the integral formulas (Appendix~\ref{sec:appendix_unconditional}) yields affine approximations for the block probabilities controlled by the first Pareto moment $\mu_1=\E[X]$ (hence requiring $\alpha>2$), for example:
\[
\begin{aligned}
P_{PP}(y)&\simeq \tfrac12+\tfrac14 y\\
P_{CP}(y)&\simeq \tfrac12+\tfrac14(y+\mu_1)\\
P_{CC}(y)&\simeq \tfrac12+\tfrac14(y+2\mu_1)
\end{aligned}
\]
At first order, motif probabilities follow by multiplying the linearized kernels; correlations induced by shared latent marks enter at higher order and are controlled by $\Var(X)$.

\paragraph*{Sparse regime and big jumps.}
As $y\to-\infty$, edges become rare unless one (or a few) core marks take atypically large values.
This is an instance of the ``big-jump'' mechanism for heavy-tailed variables \cite{Resnick2007,VezzaniBarkaiBurioni2019}.
At the level of link probabilities, the mechanism can be seen directly from the asymptotics of the logistic. For $u\ll -1$:
\[
\sigmoid(u)=\frac{1}{1+e^{-u}}
=
\frac{e^{u}}{1+e^{u}}
\sim e^{u}
\]
Hence, periphery--periphery edges have:
\[
P_{PP}(y)=\sigmoid(y)\sim e^{y}\qquad (y\to-\infty)
\]
For periphery--core edges:
\[
\begin{aligned}
P_{CP}(y)
&=
\int_{x_{\min}}^\infty C_\alpha x^{-\alpha}\sigmoid(y+x)\,dx
\sim
\\
&
\sim
\int_{|y|}^\infty C_\alpha x^{-\alpha}\,dx
=
\frac{C_\alpha}{\alpha-1}|y|^{-(\alpha-1)}
\end{aligned}
\]
where we used that $\sigmoid(y+x)\approx 1$ only when $x\gtrsim |y|$.
An analogous argument shows $P_{CC}(y)\sim |y|^{-(\alpha-1)}$ up to constants: in a core--core pair it suffices that \emph{one} of the two marks is a big jump of order $|y|$.

The same logic extends to wedges and triangles.
In symbolic labels such as CPC, CPP, or PCP the \emph{central letter identifies the central vertex} of the wedge.
As $y\to-\infty$, the dominant scaling is determined by the minimal number of distinct core marks that must overcome the negative field.
Each periphery--periphery edge contributes an additional exponential factor $e^{y}$.
The resulting asymptotic behaviors are summarized in Tables~\ref{tab:wedge_sparse} and~\ref{tab:triangle_sparse}.

\begin{table}[pos=h]
\centering
\caption{Asymptotic scaling of wedge probabilities as $y\to-\infty$.}
\label{tab:wedge_sparse}
\begin{tabular}{ll}
\toprule
Placement & $P^{\wedge}(y)$ as $y\to-\infty$ \\
\midrule
CCC & $|y|^{-(\alpha-1)}$ \\
CCP & $|y|^{-(\alpha-1)}$ \\
CPC & $|y|^{-2(\alpha-1)}$ \\
CPP & $|y|^{-(\alpha-1)}$ \\
PCP & $e^{y}\,|y|^{-(\alpha-1)}$ \\
PPP & $e^{2y}$ \\
\bottomrule
\end{tabular}
\end{table}

\begin{table}[pos=h]
\centering
\caption{Asymptotic scaling of triangle probabilities as $y\to-\infty$.}
\label{tab:triangle_sparse}
\begin{tabular}{ll}
\toprule
Placement & $P^{\triangle}(y)$ as $y\to-\infty$ \\
\midrule
CCC & $|y|^{-2(\alpha-1)}$ \\
CCP & $|y|^{-2(\alpha-1)}$ \\
CPP & $e^{y}\,|y|^{-(\alpha-1)}$ \\
PPP & $e^{3y}$ \\
\bottomrule
\end{tabular}
\end{table}

\subsection{Degree distribution: mapping marks to degrees}
Write $k_i=\sum_{j\neq i}A_{ij}$ for the degree of node $i$.
In the conditional ERGM (Sec.~\ref{sec:cp_model}), $k_i$ is Poisson--binomial with mean $\sum_{j\neq i}p_{ij}$ and variance $\sum_{j\neq i}p_{ij}(1-p_{ij})$.
In the unconditional layer, we use conditional independence and the Pareto marks to map latent heterogeneity to the degree spectrum, as in fitness (hidden-variable) models \cite{Caldarelli2002,BogunaPastorSatorras2003}.

Let $m\equiv|C|$ and $\rho\equiv m/N$.
For a \emph{core} vertex with mark $X=x$, periphery links have probability $\sigmoid(x+y)$, while core links require averaging over an i.i.d.\ Pareto mark $X'$:
\begin{equation}
\begin{aligned}
\widetilde F(x;y)
&\equiv
\E\!\left[\sigmoid(x+X'+y)\right] \\
&=
C_\alpha\int_{x_{\min}}^\infty x'^{-\alpha}\,\sigmoid(x+x'+y)\,dx'
\end{aligned}
\end{equation}
The mean degree of a marked core node is, up to $o(N)$ corrections:
\begin{equation}
k(x;y)\;=\;N\,F_\rho(x;y)
\label{eq:mean_degree_map}
\end{equation}
with \(F_\rho(x;y)\equiv (1-\rho)\,\sigmoid(x+y)+\rho\,\widetilde F(x;y)\).

Since $\sigmoid$ is increasing, $F_\rho(x;y)$ is monotone in $x$, so the map $x\mapsto q\equiv k/N$ is invertible.
As $N\to\infty$, conditional fluctuations around $k(x;y)$ are $O(\sqrt{N})$, hence degrees concentrate and the distribution follows the Jacobian rule:
\begin{equation}
\begin{aligned}
&
P(k\mid y)
\simeq
\frac{p\!\left(x(k;y)\right)}{N\,F'_\rho\!\left(x(k;y);y\right)}
\\
&
x(k;y)=F_\rho^{-1}\!\left(\frac{k}{N};y\right)
\end{aligned}
\label{eq:jacobian_degree}
\end{equation}
with:
\[
F'_\rho(x;y)
=
(1-\rho)\,\sigmoid(x+y)\bigl[1-\sigmoid(x+y)\bigr]
+\rho\,\widetilde F'(x;y)
\]
and:
\[
\widetilde F'(x;y)
=
\E\!\left[\sigmoid(x+X'+y)\bigl(1-\sigmoid(x+X'+y)\bigr)\right]
\]

\paragraph*{Dense regime ($y\to+\infty$).}
As $\sigmoid(\cdot)\to 1$ uniformly, $F_\rho(x;y)\to 1$ and degrees saturate at $N-1$:
\[
P(k\mid y\to+\infty)=\delta_{k,N-1}
\]

\paragraph*{Linear-response regime.}
When $x+x'+y=O(1)$ for typical pairs, inserting Eq.~\eqref{eq:sigmoid_linear} into Eq.~\eqref{eq:mean_degree_map} yields, for $\alpha>2$:
\begin{equation}
\begin{aligned}
&
F_\rho(x;y)\simeq
\frac12+\frac14\Bigl[x+y+\rho\,\frac{\alpha-1}{\alpha-2}\,x_{\min}\Bigr]
\\
&
F'_\rho(x;y)\simeq \frac14
\end{aligned}
\end{equation}
so that the Jacobian rule implies a Pareto shoulder in the degree distribution:
\[
P(k\mid \text{linear response})\propto k^{-\alpha}
\qquad
k_1\lesssim k\lesssim k_2
\]
with exponent matching the latent marks \cite{Caldarelli2002}.

\paragraph*{Sparse regime ($y\to-\infty$).}
Let $|y|\gg1$.
For $\rho\in(0,1]$, the periphery term in $F_\rho$ is exponentially small, while the core term scales as a Pareto tail:
heuristically, $\widetilde F(x;y)$ is dominated by those $x'$ such that $x+x'+y\gtrsim 0$, i.e., $x'\gtrsim |y|-x$, so:
\[
\widetilde F(x;y)\asymp \int_{|y|-x}^\infty x'^{-\alpha}\,dx' \asymp (|y|-x)^{1-\alpha}
\]
Therefore $F_\rho(x;y)\asymp \rho\,(|y|-x)^{1-\alpha}$.
Inverting $k=N F_\rho$ gives $|y|-x \asymp (k/N)^{-\frac{1}{\alpha-1}}$, and differentiating yields the tail exponent:
\[
P(k\mid y\ll -1)\;\propto\;\frac{1}{N}\left(\frac{k}{N}\right)^{-\frac{\alpha}{\alpha-1}},
\qquad
1\ll k\ll N
\]
Only the exponent $\frac{\alpha}{\alpha-1}$ is universal; prefactors depend on $(y,\rho,x_{\min})$.
If $\rho\to0$, the core term vanishes and degrees concentrate around $N\sigmoid(x+y)$, so no power-law tail is generated by core heterogeneity.

\paragraph*{Periphery degrees.}
Periphery vertices have no mark.
After marginalizing core marks, periphery--core links are i.i.d.\ Bernoulli with parameter $P_{CP}(y)$ and periphery--periphery links are i.i.d.\ Bernoulli with parameter $P_{PP}(y)=\sigmoid(y)$.
Hence periphery degrees are sums of binomials and are thin-tailed (asymptotically normal for $m=\Theta(N)$ by standard local CLT arguments \cite{Brandt2014DataAnalysis}).
In this model, heavy tails are a \emph{core} phenomenon.

\clearpage

\section{Empirical analysis}
\label{sec:empirical}

We apply the core--periphery ERG model of Sec.~\ref{sec:cp_model} and the PLSD correction of Sec.~\ref{sec:plsd} to two temporal datasets:
(i) weekly eMID interbank transaction networks and (ii) monthly US domestic air traffic networks.
The inference pipeline is shared, while preprocessing and scale differ across datasets.

\subsection{Common pipeline: active set, candidate cores, fitting, and validation}
\paragraph*{Active set and rolling ranking.}
For each snapshot we restrict to the \emph{active} node set (degree $>0$).
We then rank active nodes by a rolling degree score computed over the previous $B$ snapshots ($B=20$ weeks for eMID, $B=12$ months for US air) and construct candidate cores as the top-$m$ nodes for $m\in\mathcal{M}$.

\paragraph*{Model fitting and PLSD calibration.}
For each candidate size $m$ we infer parameters by minimizing either the NLL $\Phi$ (unpenalized) or the PLSD objective (penalized), both under the bound constraints $x_i\ge 0$.
In the empirical PLSD runs, the penalty scale $(\lambda_T,\lambda_W)$ is calibrated within each snapshot from the MLE scan over $m$ as described in Sec.~\ref{sec:plsd}.

\paragraph*{Core-size selection.}
To stabilize core-size selection, we use an Akaike Information Criterion (AIC) complexity penalty $2k$ with $k=1+m$ parameters (one global $y$ plus $m$ core fields) \cite{Akaike1974}.
This penalty is used \emph{only} to select $m$ and does not alter parameter estimates.
Crucially, we keep the selection rule fixed across NLL and PLSD so that differences in fit are attributable to the PLSD correction rather than to model-selection artifacts.

\paragraph*{Model checking.}
For each snapshot and fitted parameter set, we generate Monte Carlo networks from the fitted dyad-independent model and compute $L$, $W$, and $T$.
We summarize agreement both through time-series plots (empirical count vs simulated mean with dispersion bands) and through a cross-snapshot normalized RMSE scoreboard.
For a motif count $M\in\{L,W,T\}$, define:
\[
\mathrm{nRMSE}_M
=
\sqrt{
\frac{1}{|\mathcal{T}|}
\sum_{t\in\mathcal{T}}
\left(
\frac{\E_\theta[M_t]-M_{t,\mathrm{obs}}}{M_{t,\mathrm{obs}}}
\right)^2
}
\]
where $\mathcal{T}$ is the set of snapshots with $M_{t,\mathrm{obs}}>0$ and $\E_\theta[M_t]$ is estimated by Monte Carlo averages.
(We report relative errors because scales differ widely across datasets and time.)

Beyond motif counts, we also report network-level diagnostics and the parameter shift induced by PLSD on the selected core fields.
For a node belonging to both the NLL-selected and PLSD-selected cores in the same snapshot, define:
\[
\Delta x_i=x_{i,\mathrm{PLSD}}-x_{i,\mathrm{NLL}}
\]
The empirical distribution of $\Delta x_i$ shows whether the structural correction acts approximately uniformly across the common core or is concentrated on a subset of highly active core nodes.
Tables~\ref{tab:network_metrics_scoreboard}--\ref{tab:core_detection_summary} and Fig.~\ref{fig:delta_x_common_core} summarize the cross-dataset diagnostics used to compare motif fit and core selection.

\begin{center}
\centering
\captionof{table}{Out-of-sample network-level diagnostics for the empirical fits. Errors for $L$, $W$, $T$, $C$, $r$, and $Q$ are relative and reported in percent; ASPL and diameter are reported as signed biases.}
\label{tab:network_metrics_scoreboard}
\begin{tabular}{llrrrrrrrr}
\toprule
Dataset & Criterion & $L$ & $W$ & $T$ & $C$ & $r$ & $Q$ & ASPL & Diam. \\
\midrule
eMID & NLL + AIC & 0.147 & 3.83 & 109.5 & 9.68 & 6.92 & 5.35 & -0.15 & -1.26 \\
eMID & PLSD + AIC & 20.43 & 0.345 & 35.42 & 1.70 & 14.58 & 19.40 & -0.29 & -1.93 \\
US air & NLL + AIC & 0.031 & 0.661 & 15.21 & 7.90 & 19.35 & 4.92 & -0.16 & -1.22 \\
US air & PLSD + AIC & 3.28 & 0.049 & 0.094 & 0.036 & 18.40 & 2.37 & -0.11 & -1.18 \\
\bottomrule
\end{tabular}
\end{center}

\clearpage
\begin{center}
\centering
\captionof{table}{Core-detection comparison between maximum likelihood and PLSD. $\Delta m=m_{\mathrm{PLSD}}-m_{\mathrm{NLL}}$; Jaccard compares the selected core sets; $\Delta x_i=x_{i,\mathrm{PLSD}}-x_{i,\mathrm{NLL}}$ is computed on nodes belonging to both selected cores in the same snapshot.}
\label{tab:core_detection_summary}
\begin{tabular}{lrrrrrrr}
\toprule
Dataset & Snapshots & med. $m_{\mathrm{NLL}}$ & med. $m_{\mathrm{PLSD}}$ & med. $\Delta m$ & med. Jaccard & med. $\Delta x_i$ & 10--90\% $\Delta x_i$ \\
\midrule
eMID & 272 & 41.0 & 21.5 & -16.0 & 0.55 & -0.88 & [-2.00, 0.09] \\
US air & 120 & 196.5 & 198.5 & 2.5 & 0.98 & 0.38 & [-1.17, 1.41] \\
\bottomrule
\end{tabular}
\end{center}

\begin{center}
\begin{minipage}{0.49\columnwidth}
\centering
\includegraphics[width=\linewidth]{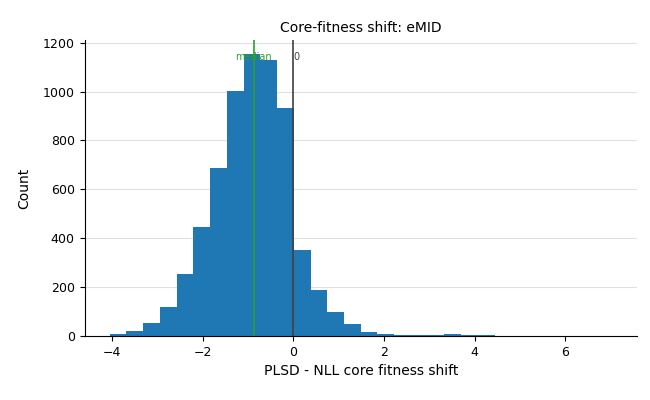}
\par\smallskip{\small (a) eMID.}
\end{minipage}
\begin{minipage}{0.49\columnwidth}
\centering
\includegraphics[width=\linewidth]{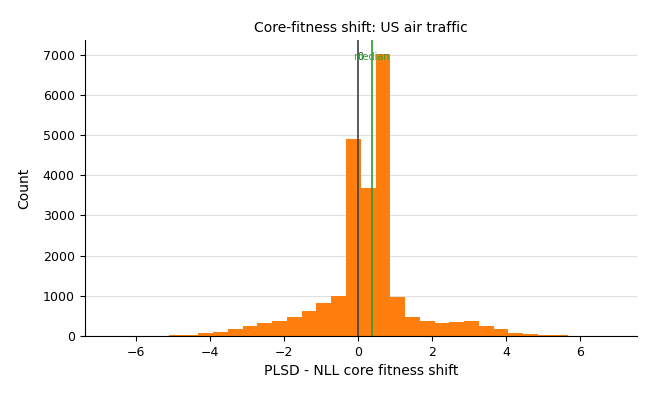}
\par\smallskip{\small (b) US air traffic.}
\end{minipage}
\captionof{figure}{Distribution of PLSD-induced shifts in core fitness, $\Delta x_i$, for nodes selected in the core by both NLL and PLSD in the same snapshot. The vertical black line marks zero and the green line marks the median shift.}
\label{fig:delta_x_common_core}
\end{center}

The PLSD correction improves not only the targeted wedge and triangle counts, but also several network-level diagnostics, most notably clustering in both datasets.
The effect on the selected core is dataset-dependent rather than mechanically monotone.
In eMID, PLSD selects smaller but structurally more motif-consistent cores, with a negative median shift in common-core fitness.
In US air traffic, the selected core size changes little and the overlap between NLL and PLSD cores remains high, while the positive median $\Delta x_i$ indicates a reweighting of activity inside an already stable core.
Thus PLSD uses triangle and wedge information to reallocate core intensity and, depending on the empirical network, may change either the size or the internal heterogeneity of the core.

\subsection{eMID: weekly interbank networks (2009--2015)}
\paragraph*{Data and preprocessing.}
We use credit transactions between banks operating in the electronic Market for Interbank Deposit (eMID) from 2009-10-19 to 2015-01-02.
Daily directed transaction records are aggregated into 292 weekly binary snapshots.
For week $t$, the adjacency matrix has entry $A_{t,ij}=1$ if at least one transaction between banks $i$ and $j$ is observed within the week and $A_{t,ij}=0$ otherwise.
We then construct undirected, unweighted graphs by logical OR aggregation: an undirected edge is present whenever at least one directed transaction is observed in either direction during the weekly window; self-loops are removed.
Our empirical object is therefore a weekly, undirected, binary projection rather than the model-comparison setting considered by Barucca and Lillo \cite{BaruccaLillo2016}. Their eMID results nevertheless make the methodological issue directly relevant: the inferred bipartite or core--periphery organization depends on degree correction and on temporal aggregation. Here we hold the observation rule and dyad-independent model family fixed and ask whether a motif-aware penalty can reduce the remaining structural misspecification.
The active node set varies over time (56--109 active banks per week, median 73), and the AIC-selected core size ranges from 4 to 77 with a median of 22 \cite{IazzettaManna2009,FrickeLux2015,BaruccaLillo2018}.

\begin{figure}[pos=H]
\centering
\begin{subfigure}{0.49\columnwidth}
\centering
\includegraphics[width=\linewidth]{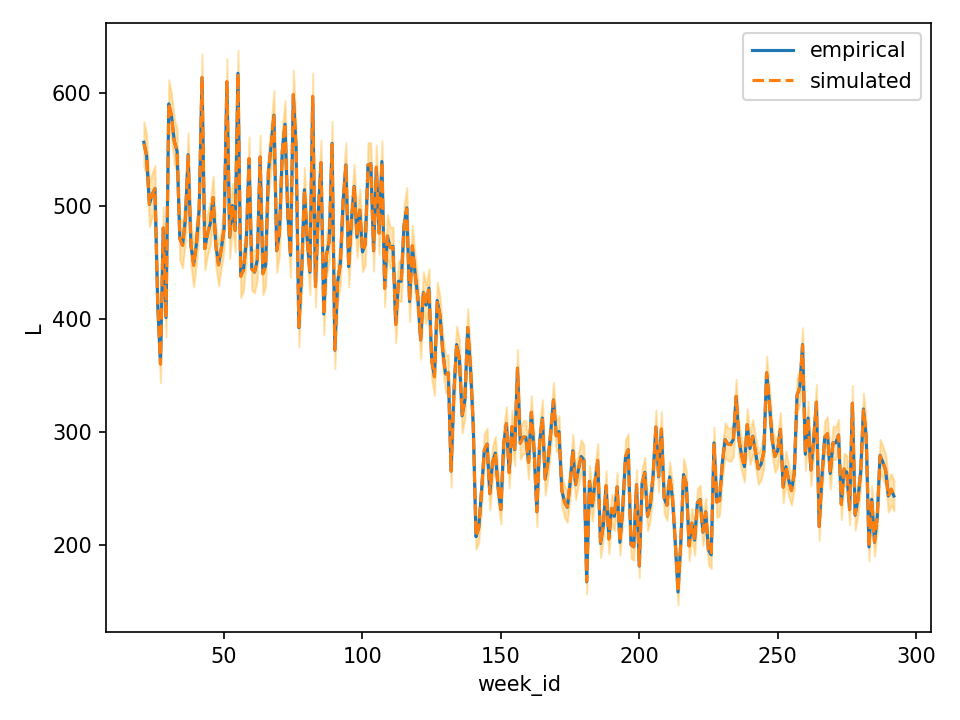}
\caption{$L$ series (NLL + AIC).}
\end{subfigure}
\vspace{0.3em}
\begin{subfigure}{0.49\columnwidth}
\centering
\includegraphics[width=\linewidth]{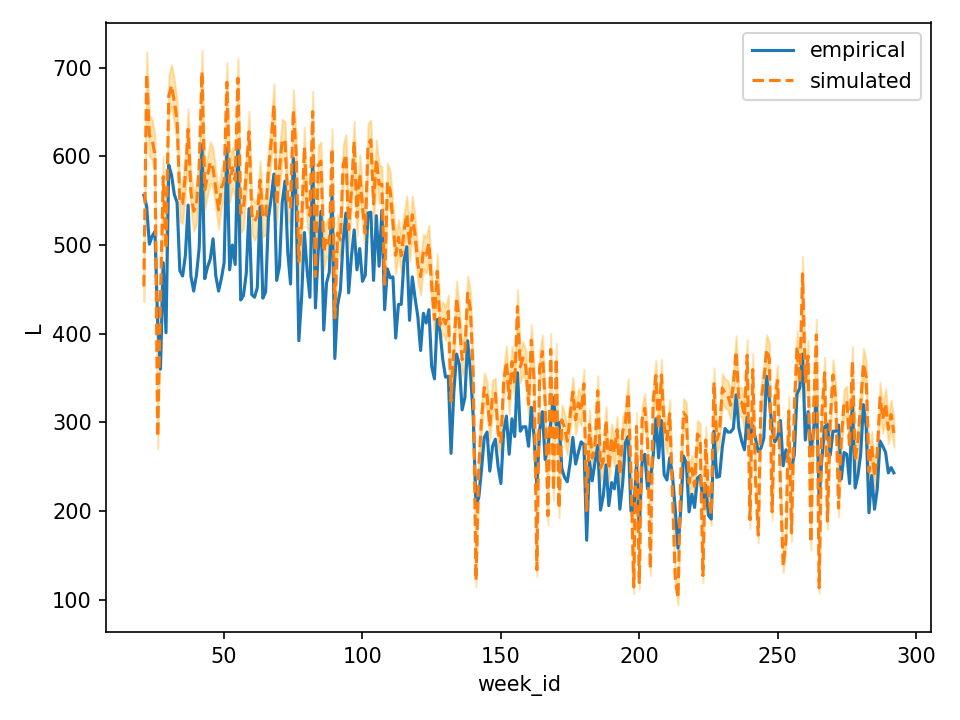}
\caption{$L$ series (PLSD + AIC).}
\end{subfigure}
\caption{eMID link series: empirical counts versus Monte Carlo means under the fitted models.}
\label{fig:emid_L}
\end{figure}

\begin{figure}[pos=H]
\centering
\begin{subfigure}{0.49\columnwidth}
\centering
\includegraphics[width=\linewidth]{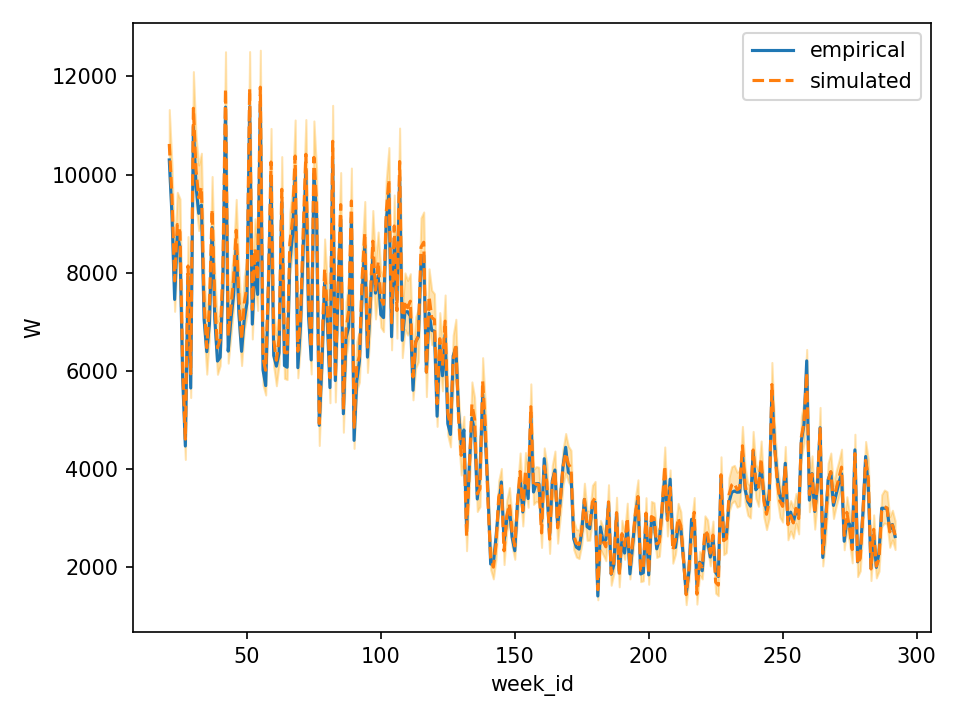}
\caption{$W$ series (NLL + AIC).}
\end{subfigure}
\vspace{0.3em}
\begin{subfigure}{0.49\columnwidth}
\centering
\includegraphics[width=\linewidth]{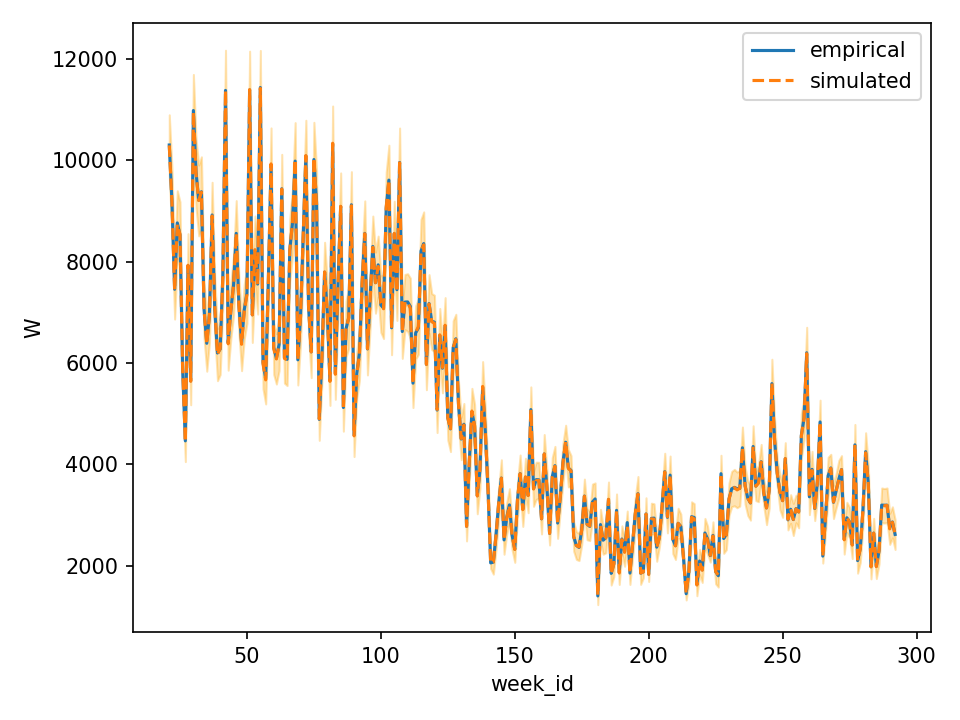}
\caption{$W$ series (PLSD + AIC).}
\end{subfigure}
\caption{eMID wedge series.}
\label{fig:emid_W}
\end{figure}

\begin{figure}[pos=H]
\centering
\begin{subfigure}{0.49\columnwidth}
\centering
\includegraphics[width=\linewidth]{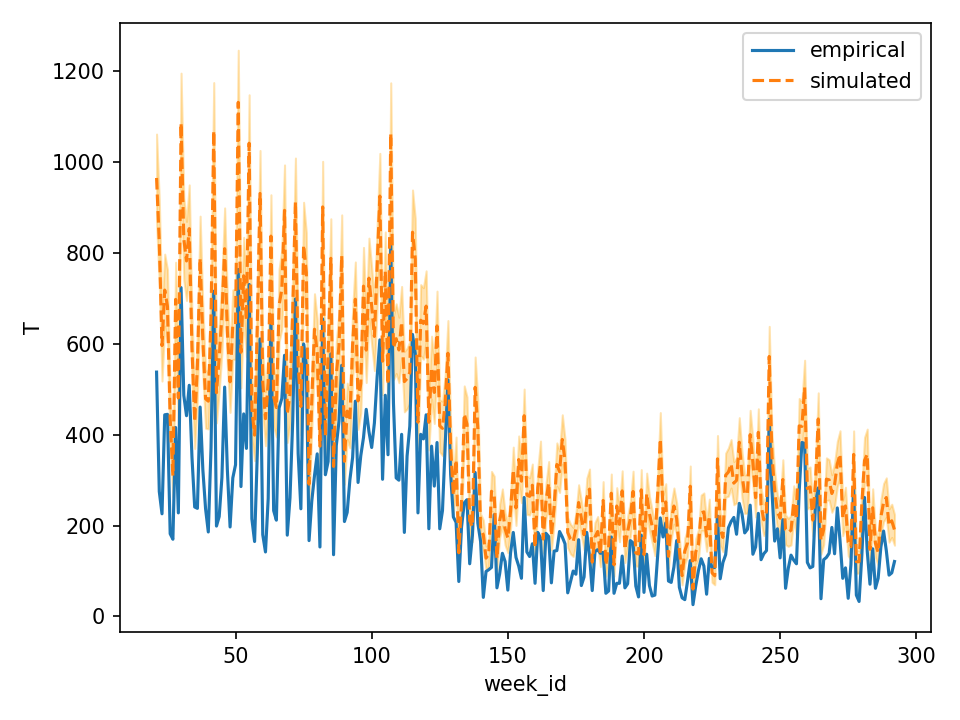}
\caption{$T$ series (NLL + AIC).}
\end{subfigure}
\vspace{0.3em}
\begin{subfigure}{0.49\columnwidth}
\centering
\includegraphics[width=\linewidth]{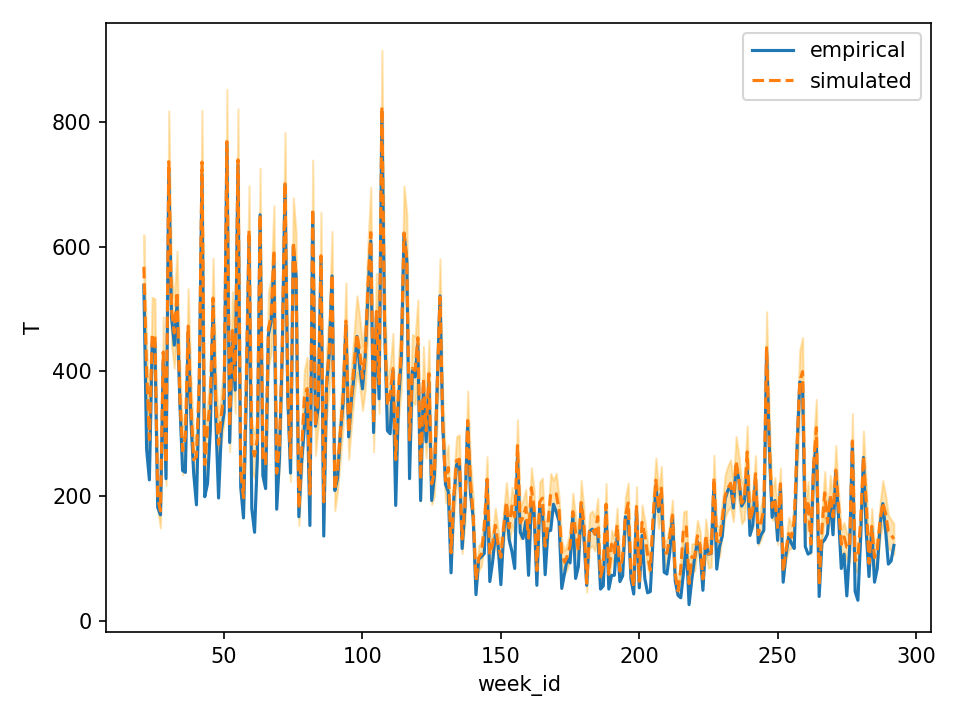}
\caption{$T$ series (PLSD + AIC).}
\end{subfigure}
\caption{eMID triangle series.}
\label{fig:emid_T}
\end{figure}

\paragraph*{Results.}
Table~\ref{tab:emid_scoreboard} summarizes the PLSD trade-off between link fit and higher-order motifs.
Under NLL, links are matched extremely well ($n\mathrm{RMSE}_L\approx0.14\%$) but wedges and triangles remain substantially misfit.
PLSD rebalances the fit toward motifs, reducing wedge discrepancies by more than an order of magnitude and substantially decreasing triangle mismatch, at the expense of larger link errors.
Figures~\ref{fig:emid_L}--\ref{fig:emid_T} display the corresponding time series.

\begin{table}[pos=H]
\centering
\caption{eMID weekly motif-fit scoreboard.}
\label{tab:emid_scoreboard}
\small
\begin{tabular}{@{}lccc@{}}
\toprule
Configuration & nRMSE$_L$ (\%) & nRMSE$_W$ (\%) & nRMSE$_T$ (\%) \\
\midrule
NLL & 0.14\% & 5.80\% & 144.56\% \\
NLL+AIC & 0.15\% & 3.83\% & 109.53\% \\
PLSD & 20.22\% & 0.32\% & 36.78\% \\
PLSD+AIC & 20.43\% & 0.35\% & 35.42\% \\
\bottomrule
\end{tabular}
\end{table}

\FloatBarrier

\subsection{US air traffic: monthly networks (1991--2000)}
\paragraph*{Data and preprocessing.}
Monthly US air traffic networks are derived from domestic carrier statistics over 1991-01-01 to 2000-12-31 (132 months) \cite{BTS_T100}.
For each month, we start from directed airport-pair records and retain a binary connection whenever at least one domestic service is observed between the two airports during that month.
Directed, unweighted edge lists are converted to undirected, unweighted graphs by the same logical OR rule used for eMID: an undirected edge is present if traffic is observed in at least one direction; self-loops are removed.
The active node set ranges from 293 to 421 airports per month (median 342), and the AIC-selected core size ranges from 168 to 246 with a median of 199, reflecting a substantially larger and denser network than eMID.

\begin{figure}[pos=H]
\centering
\begin{subfigure}{0.49\columnwidth}
\centering
\includegraphics[width=\linewidth]{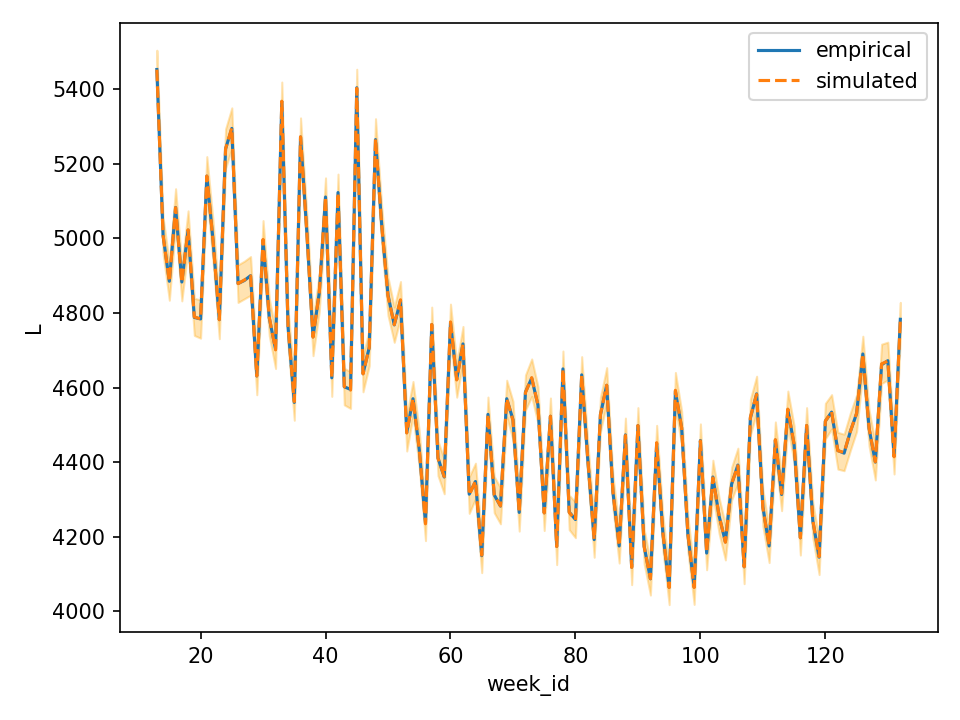}
\caption{$L$ series (NLL + AIC).}
\end{subfigure}
\vspace{0.3em}
\begin{subfigure}{0.49\columnwidth}
\centering
\includegraphics[width=\linewidth]{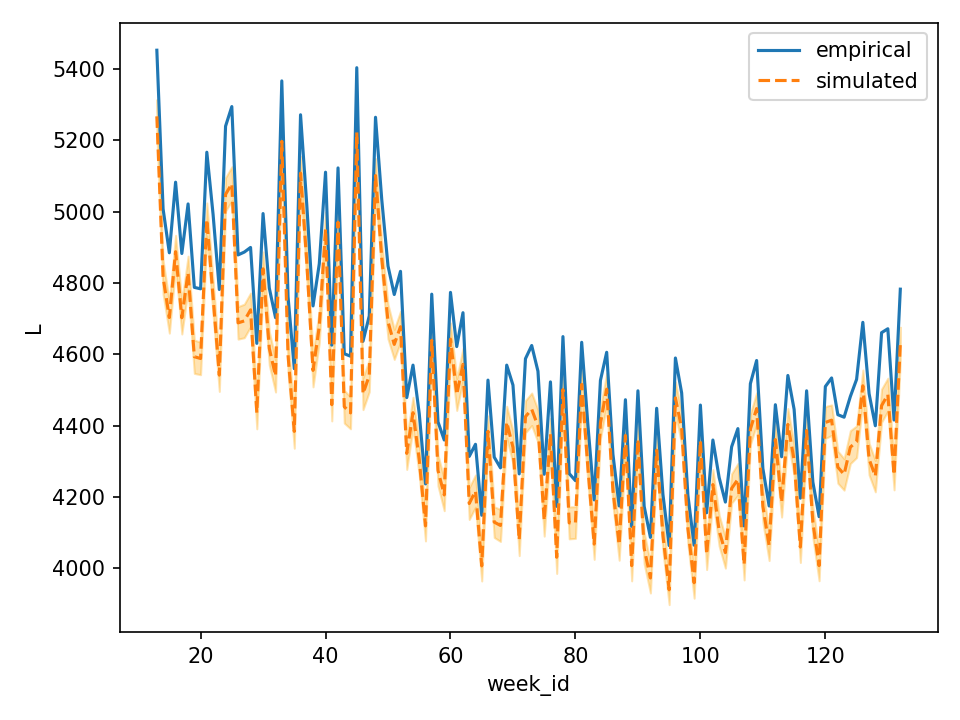}
\caption{$L$ series (PLSD + AIC).}
\end{subfigure}
\caption{US air traffic link series.}
\label{fig:usair_L}
\end{figure}

\begin{figure}[pos=H]
\centering
\begin{subfigure}{0.49\columnwidth}
\centering
\includegraphics[width=\linewidth]{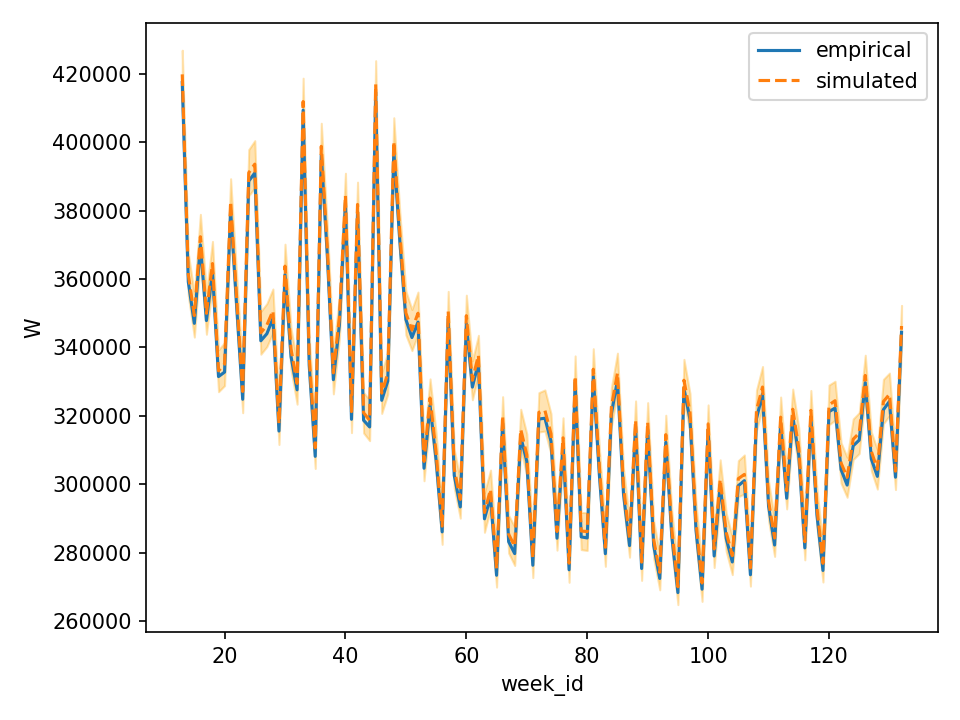}
\caption{$W$ series (NLL + AIC).}
\end{subfigure}
\vspace{0.3em}
\begin{subfigure}{0.49\columnwidth}
\centering
\includegraphics[width=\linewidth]{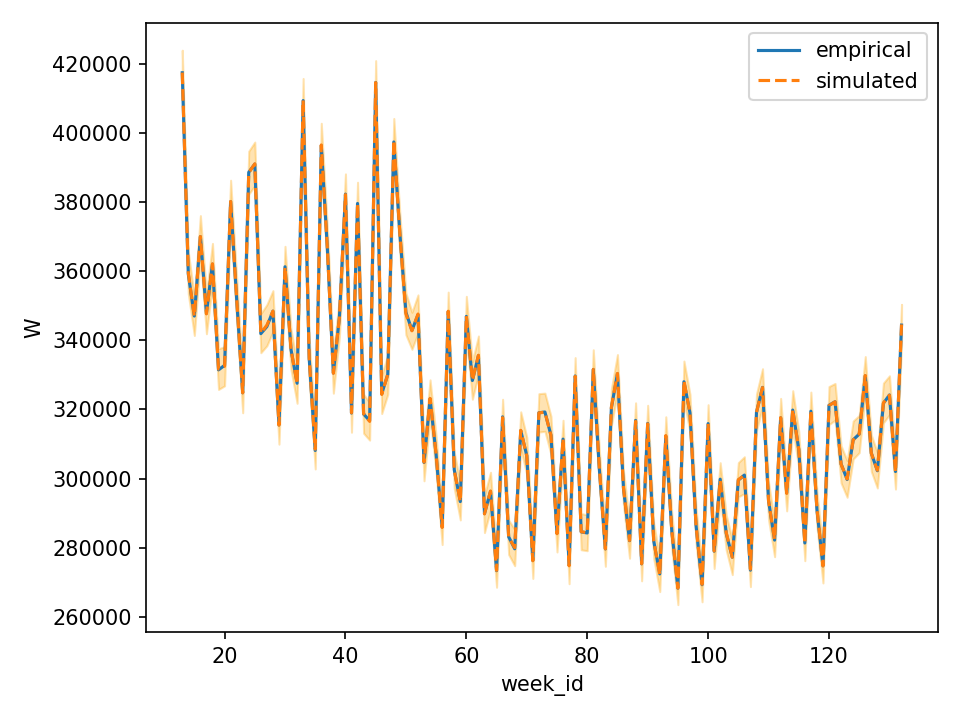}
\caption{$W$ series (PLSD + AIC).}
\end{subfigure}
\caption{US air traffic wedge series.}
\label{fig:usair_W}
\end{figure}

\begin{figure}[pos=H]
\centering
\begin{subfigure}{0.49\columnwidth}
\centering
\includegraphics[width=\linewidth]{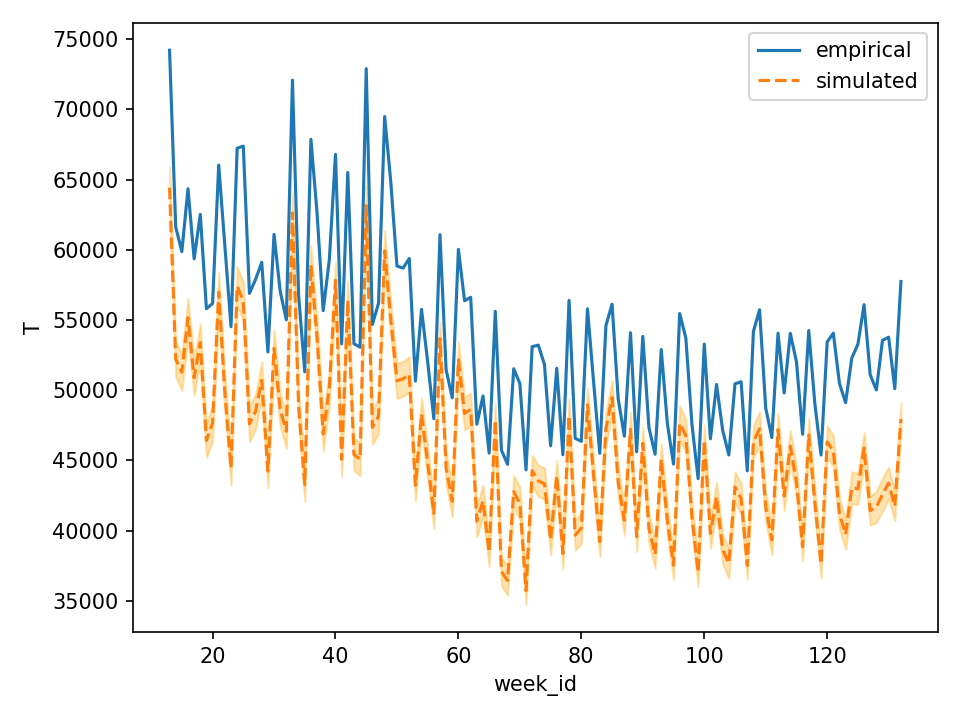}
\caption{$T$ series (NLL + AIC).}
\end{subfigure}
\vspace{0.3em}
\begin{subfigure}{0.49\columnwidth}
\centering
\includegraphics[width=\linewidth]{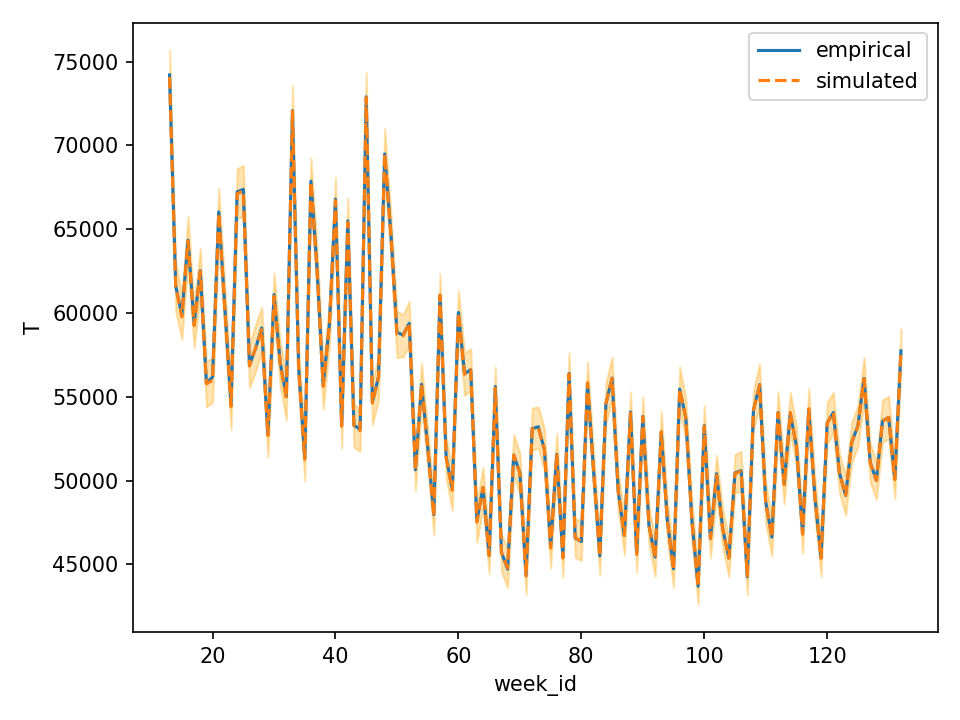}
\caption{$T$ series (PLSD + AIC).}
\end{subfigure}
\caption{US air traffic triangle series.}
\label{fig:usair_T}
\end{figure}

\paragraph*{Results.}
Table~\ref{tab:usair_scoreboard} summarizes the PLSD trade-off between link fit and higher-order motifs.
Under NLL, links are matched extremely well ($n\mathrm{RMSE}_L\approx0.03\%$) but wedges and triangles remain misfit, especially triangles ($\approx15\%$).
PLSD rebalances the fit toward motifs, reducing wedge and triangle errors by over an order of magnitude to $\approx0.05\%$ and $\approx0.095\%$, at the expense of a few percent link error.
Figures~\ref{fig:usair_L}--\ref{fig:usair_T} display the corresponding time series.

\begin{center}
\centering
\captionof{table}{US Air Traffic monthly motif-fit scoreboard.}
\label{tab:usair_scoreboard}
\small
\begin{tabular}{@{}lccc@{}}
\toprule
Configuration & nRMSE$_L$ (\%) & nRMSE$_W$ (\%) & nRMSE$_T$ (\%) \\
\midrule
NLL & 0.0347\% & 0.7439\% & 14.8\% \\
NLL+AIC & 0.0306\% & 0.6612\% & 15.21\% \\
PLSD & 3.309\% & 0.0560\% & 0.0955\% \\
PLSD+AIC & 3.277\% & 0.0491\% & 0.0943\% \\
\bottomrule
\end{tabular}
\end{center}
\FloatBarrier

\paragraph*{Cross-dataset comparison.}
Across both datasets, PLSD consistently trades link accuracy for motif agreement.
The magnitude of the trade-off is dataset dependent: in eMID, improving triangles and wedges substantially increases link error, while in US air traffic the same penalty yields sharp motif gains at only a few percent link cost.
This difference reflects that the unpenalized model starts closer to the US air motifs (smaller baseline wedge/triangle errors), so PLSD has less discrepancy to correct there than in eMID.
Equivalently, the ``distance'' between the dyad-independent null and the empirical graph is smaller for US air, consistent with domain-specific expressions of triadic closure and heterogeneity across scales.

The calibrated penalty scales show the same cross-dataset contrast.
Table~\ref{tab:lambda_eff_summary} reports $\lambda_{\mathrm{eff}}$ summarized across snapshots for the four empirical runs using PLSD.
Effective penalties are about two orders of magnitude larger in eMID than in US air, indicating that the motif discrepancies are structurally larger in eMID and require stronger correction to reach a comparable likelihood--motif balance.
Within each dataset, the AIC-selected cores yield slightly smaller $\lambda_{\mathrm{eff}}$ than the fixed-core PLSD runs, suggesting that model selection absorbs part of the discrepancy that would otherwise be corrected by the penalty.
These values therefore quantify, in a normalized scalar, how far each dataset sits from the dyad-independent core--periphery baseline.

\begin{table}[pos=H]
\centering
\caption{Summary of empirical $\lambda_{\mathrm{eff}}$ across snapshots (mean with s.d., and median).}
\label{tab:lambda_eff_summary}
\small
\begin{tabular}{lcc}
\toprule
Run & Mean (s.d.) & Median \\
\midrule
eMID PLSD & $3.51\times10^{-2}\,(6.22\times10^{-2})$ & $1.22\times10^{-2}$ \\
eMID PLSD + AIC & $2.90\times10^{-2}\,(6.03\times10^{-2})$ & $4.00\times10^{-3}$ \\
US air PLSD & $1.00\times10^{-4}\,(1.51\times10^{-4})$ & $3.99\times10^{-5}$ \\
US air PLSD + AIC & $7.86\times10^{-5}\,(1.18\times10^{-4})$ & $2.69\times10^{-5}$ \\
\bottomrule
\end{tabular}
\end{table}

\clearpage

\section{Conclusions and outlook}
\label{sec:conclusions}

PLSD provides a tractable way to correct structural misspecification at the level of inference.
By augmenting a dyad-independent likelihood with a standardized motif discrepancy penalty, the method directly targets higher-order structure (here wedges and triangles) while retaining a likelihood anchor and a meaningful parameter interpretation.
In simulations, PLSD behaves as expected: it is inert under correct specification and becomes active under targeted misspecifications, delivering large motif improvements at small predictive cost.
In a core--periphery setting, we complement the inference layer with an unconditional Pareto-marked layer that yields semi-closed motif formulas and a clear sparse-regime scaling picture via big jumps.
Empirically, the method transfers across domains and scales, substantially improving motif fit on eMID and US air traffic networks, with dataset-dependent link-level costs.
This contribution is complementary to model-selection analyses in which core--periphery structure changes with the null model or temporal aggregation \cite{BaruccaLillo2016}: PLSD instead holds a tractable model family fixed, targets its higher-order discrepancies, and makes the associated likelihood trade-off explicit.

\paragraph*{Outlook.}
Natural extensions include directed and weighted networks, time-adaptive penalties, and penalties targeting larger motif families or block-specific closure.
A further direction is to connect PLSD more explicitly to minimum-distance estimation and generalized method-of-moments theory (e.g., optimal weighting of correlated discrepancies).

\section*{GitHub repository}
The code and configuration to reproduce our findings are available in the \href{https://github.com/antoniomosca27/core-periphery-us-air-plsd}{project's GitHub repository}. This repository includes the US domestic air traffic dataset, which is publicly available from \href{https://networks.skewed.de/net/us_air_traffic}{Netzschleuder} and was originally obtained from the Bureau of Transportation Statistics. The eMID interbank dataset is not shared in the repository as these data are private.

\clearpage

\appendix

\section{Additional misspecification: heterogeneous core fitness}
\label{sec:appendix_hetero}
\begin{figure}[pos=H]
\centering
\begin{subfigure}{0.80\columnwidth}
\centering
\includegraphics[width=\linewidth]{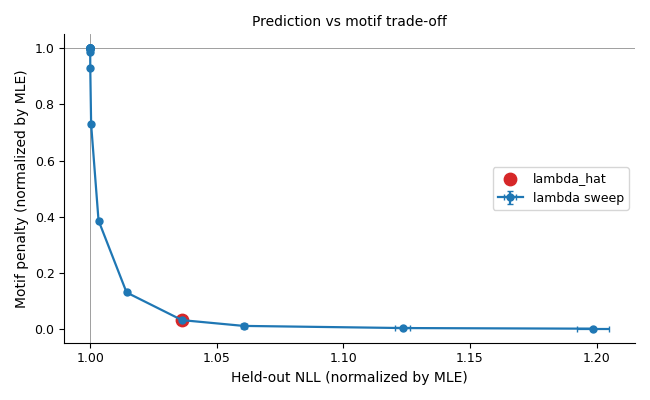}
\caption{Trade-off curve (heterogeneous core).}
\end{subfigure}
\vspace{0.4em}
\begin{subfigure}{0.95\columnwidth}
\centering
\includegraphics[width=\linewidth]{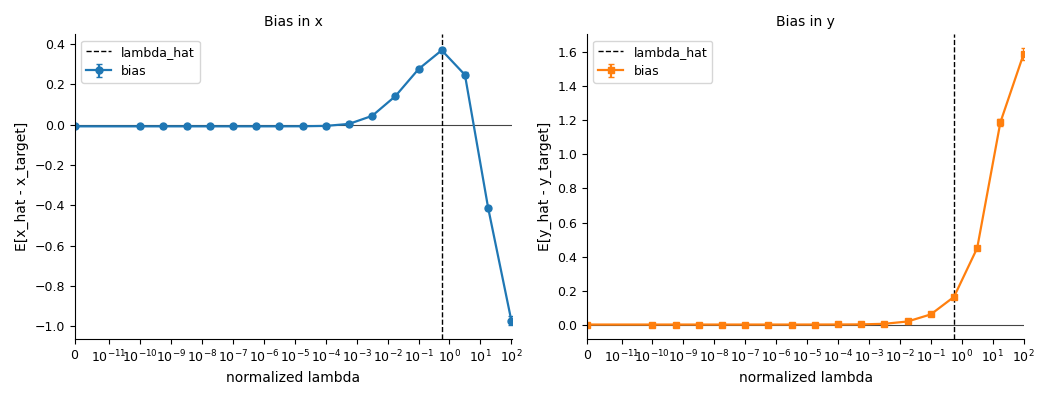}
\caption{Bias vs $\lambda$ (heterogeneous core).}
\end{subfigure}
\caption{Trade-off curve and parameter biases under PLSD for heterogeneous case.}
\label{fig:toy_hetero}
\end{figure}

We generate data from the core--periphery model introduced above with heterogeneous core fitnesses: $X_i\sim\mathrm{Pareto}(\alpha=2.5, x_{\min}=0.72)$ and $y^*=-5.0$.
We then fit the data with the usual toy (flat-core) model, so the misspecification comes from the data--fit mismatch rather than from adding triangles as in Sec.~\ref{sec:toy}.
Table~\ref{tab:pareto_stats} summarizes the fitness distribution, with parameters chosen to match the Pareto mean with $x^*$.
PLSD selects $\hat\lambda=0.562$, reducing motif mismatch at a modest increase in held-out $\NLL$ (Fig.~\ref{fig:toy_hetero}).

\begin{center}
\centering
\captionof{table}{Pareto core fitness summaries (mean across replicates).}
\label{tab:pareto_stats}
\begin{tabular}{lcccc}
\hline
Mean & 5th pct & Median & 95th pct \\
\hline
1.194 & 0.616 & 0.850 & 2.649 \\
\hline
\end{tabular}
\end{center}

Table~\ref{tab:simulation_baseline_vs_hetero} reports the corresponding held-out selection summary.

\begin{center}
\captionof{table}{Balanced $\lambda$ selection summary for the toy model. Normalized metrics and $J(\hat\lambda)$ are evaluated on held-out graphs; values are mean (SE) over replicates. $\text{NLL}_{\text{norm}}$ and $\text{Zpen}_{\text{norm}}$ are normalized by their $\lambda=0$ test values.}
\label{tab:simulation_baseline_vs_hetero}
\small
\begin{tabular}{lcc}
\toprule
Metric & Baseline & Misspecified \\
\midrule
$\hat\lambda$ (norm.) & 1.78e-08 & 0.562 \\
$\text{NLL}_{\text{norm}}(\hat\lambda)$ & 1.000 (0.000) & 1.036 (0.001) \\
$\text{Zpen}_{\text{norm}}(\hat\lambda)$ & 1.000 (0.001) & 0.031 (0.001) \\
$J(\hat\lambda)$ & 1.000 (0.001) & 0.534 (0.001) \\
$\hat x$ & 1.198 (0.001) & 1.856 (0.010) \\
$\hat y$ & -4.999 (0.001) & -4.714 (0.006) \\
\bottomrule
\end{tabular}
\end{center}


\FloatBarrier
\clearpage
\section{Additional simulation diagnostics}
\label{sec:appendix_diagnostics}
\begin{figure}
\centering
\begin{subfigure}{0.48\columnwidth}
\centering
\includegraphics[width=\linewidth]{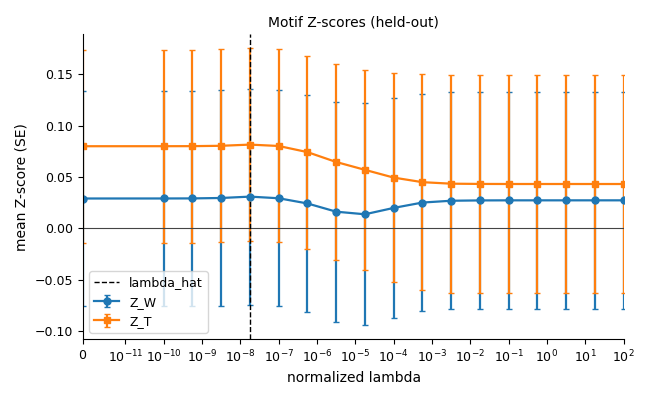}
\caption{Z-scores.}
\end{subfigure}
\begin{subfigure}{0.48\columnwidth}
\centering
\includegraphics[width=\linewidth]{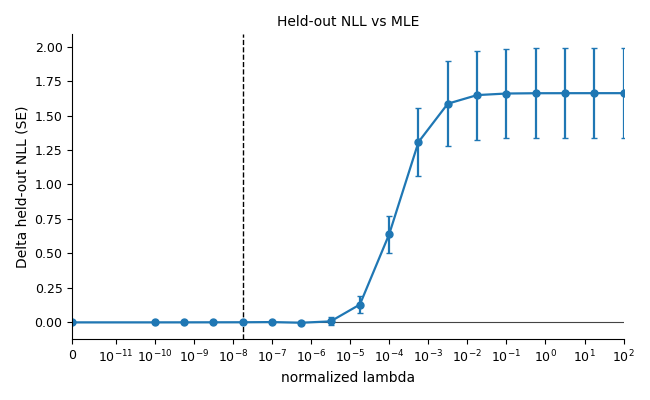}
\caption{Held-out NLL.}
\end{subfigure}
\caption{Motif Z-scores and held-out NLL costs vs $\lambda$ (baseline).}
\label{fig:zscores_nll_baseline}
\end{figure}

\begin{figure}
\centering
\begin{subfigure}{0.48\columnwidth}
\centering
\includegraphics[width=\linewidth]{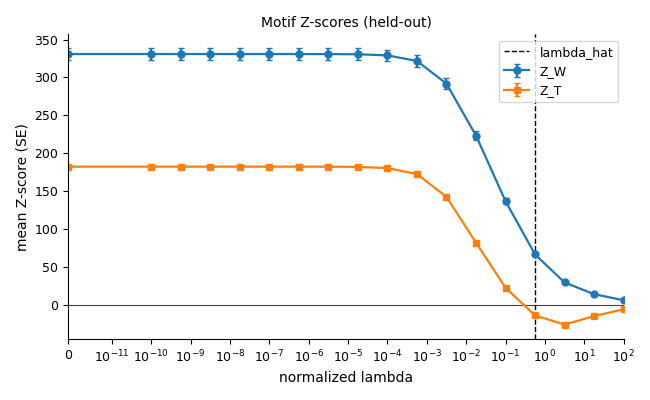}
\caption{Z-scores.}
\end{subfigure}
\begin{subfigure}{0.48\columnwidth}
\centering
\includegraphics[width=\linewidth]{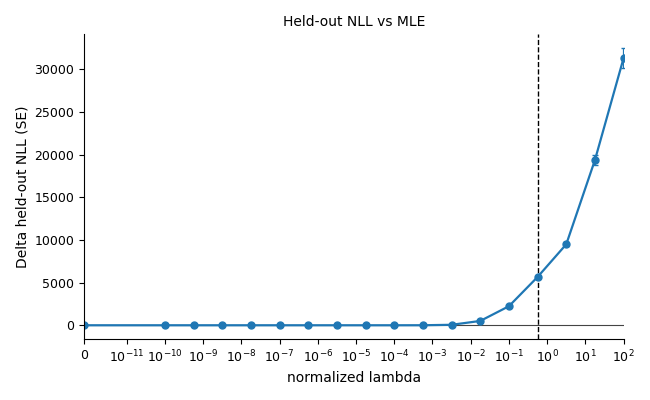}
\caption{Held-out NLL.}
\end{subfigure}
\caption{Motif Z-scores and held-out NLL costs vs $\lambda$ (heterogeneous core).}
\label{fig:zscores_nll_hetero}
\end{figure}

\begin{figure}
\centering
\begin{subfigure}{0.48\columnwidth}
\centering
\includegraphics[width=\linewidth]{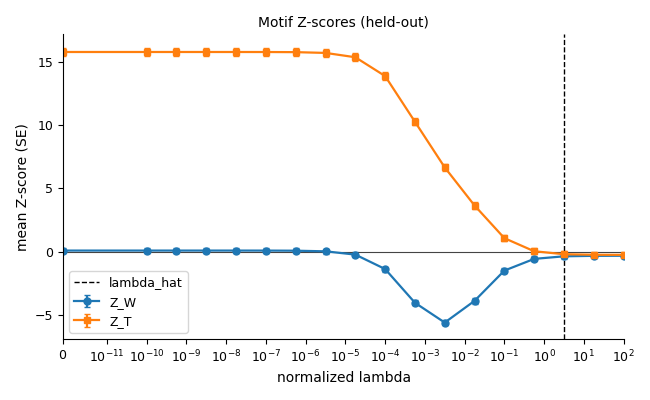}
\caption{Z-scores.}
\end{subfigure}
\begin{subfigure}{0.48\columnwidth}
\centering
\includegraphics[width=\linewidth]{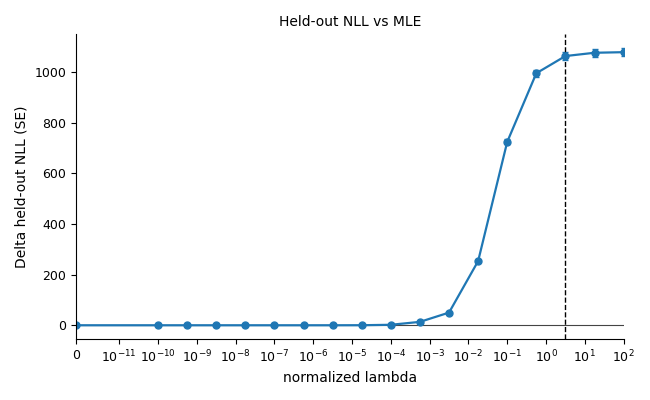}
\caption{Held-out NLL.}
\end{subfigure}
\caption{Motif Z-scores and held-out NLL costs vs $\lambda$ (triadic closure).}
\label{fig:zscores_nll_triadic}
\end{figure}

Figures~\ref{fig:varmse_baseline}--\ref{fig:varmse_triadic} and \ref{fig:zscores_nll_baseline}--\ref{fig:zscores_nll_triadic} report variance/MSE curves, Z-scores versus $\lambda$, and held-out NLL costs for the toy simulations.

\begin{center}
\centering
\includegraphics[width=0.82\columnwidth]{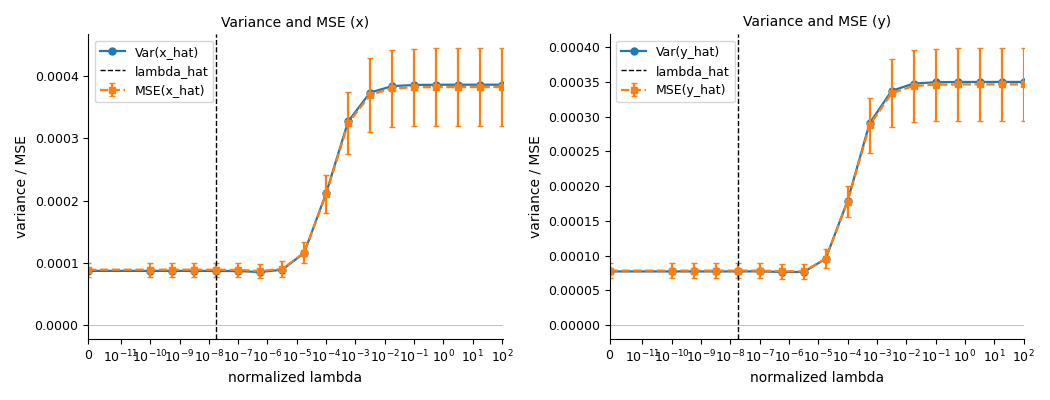}
\captionof{figure}{Variance and MSE vs $\lambda$: baseline.}
\label{fig:varmse_baseline}
\end{center}

\begin{center}
\centering
\includegraphics[width=0.82\columnwidth]{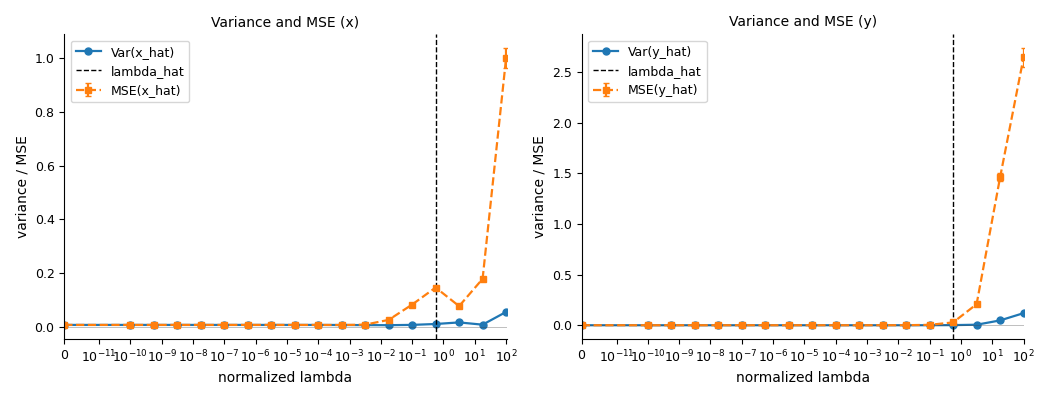}
\captionof{figure}{Variance and MSE vs $\lambda$: heterogeneous core.}
\label{fig:varmse_hetero}
\end{center}

\begin{center}
\centering
\includegraphics[width=0.82\columnwidth]{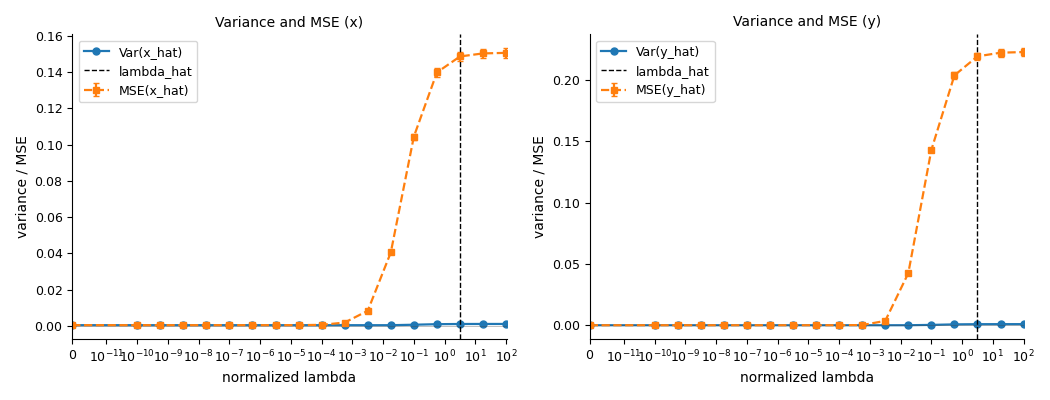}
\captionof{figure}{Variance and MSE vs $\lambda$: triadic closure.}
\label{fig:varmse_triadic}
\end{center}


\section{Small-$\lambda$ linear response derivation}
\label{sec:appendix_linear}
Let:
\[
\Phi_\lambda(\theta)=\Phi(\theta)+\lambda\,\frac{S_\Phi}{S_Z}\,Q(\theta),
\quad
Q(\theta)=Z_W(\theta)^2+Z_T(\theta)^2
\]
where the scale factors $S_\Phi=\Phi(\hat\theta_0)$ and $S_Z=Z_W(\hat\theta_0)^2+Z_T(\hat\theta_0)^2$ are evaluated at the unpenalized MLE $\hat\theta_0$ and then held fixed.
Assume $\Phi$ is twice continuously differentiable and its Hessian $H_0=\nabla^2\Phi(\hat\theta_0)$ is nonsingular.
The first-order condition can be written as:
\[
F(\theta,\lambda)=\nabla\Phi(\theta)+\lambda\,\frac{S_\Phi}{S_Z}\,\nabla Q(\theta)=0
\]
By the implicit function theorem, there exists a differentiable curve $\theta(\lambda)$ with $\theta(0)=\hat\theta_0$ solving $F(\theta(\lambda),\lambda)=0$ for small $\lambda$.
Differentiating at $\lambda=0$ yields:
\[
H_0\,\theta'(0)+\frac{S_\Phi}{S_Z}\,g_0=0,
\qquad
g_0=\nabla Q(\hat\theta_0)
\]
so:
\[
\theta'(0)=-\frac{S_\Phi}{S_Z}\,H_0^{-1}g_0
\]
Equivalently, writing $\Delta\theta=\hat\theta_\lambda-\hat\theta_0$ and expanding:
\[
\begin{aligned}
\nabla\Phi(\hat\theta_0+\Delta\theta)
=
H_0\,\Delta\theta + \mathcal{O}(\|\Delta\theta\|^2)
\\
\nabla Q(\hat\theta_0+\Delta\theta)=g_0+\mathcal{O}(\|\Delta\theta\|)
\end{aligned}
\]
gives:
\[
H_0\,\Delta\theta+\lambda\,\frac{S_\Phi}{S_Z}\,g_0=\mathcal{O}(\|\Delta\theta\|^2+\lambda\|\Delta\theta\|)
\]
and therefore:
\[
\Delta\theta=-\lambda\,\frac{S_\Phi}{S_Z}\,H_0^{-1}g_0+\mathcal{O}(\lambda^2)
\]
which is Eq.~(\ref{eq:linear_response}).

\begin{figure}
\centering
\includegraphics[width=0.95\columnwidth]{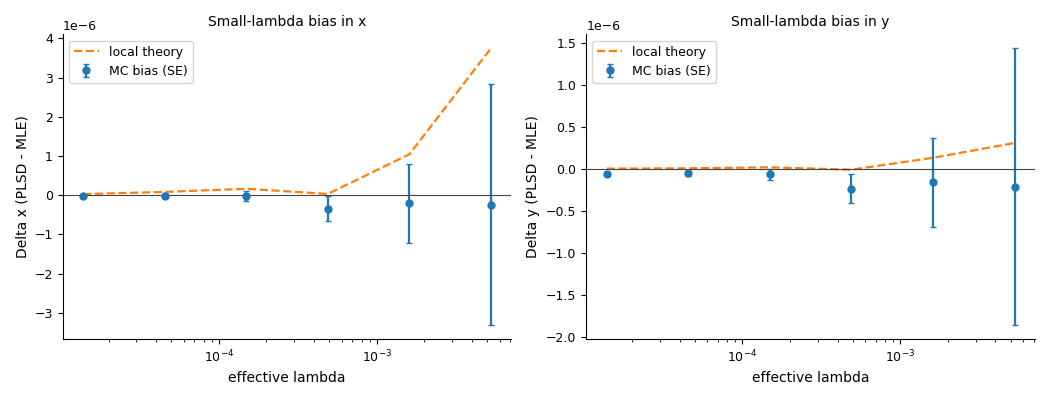}
\caption{Small-$\lambda$ linear response: baseline case.}
\label{fig:smalllambda_baseline}
\end{figure}

\begin{figure}
\centering
\includegraphics[width=0.95\columnwidth]{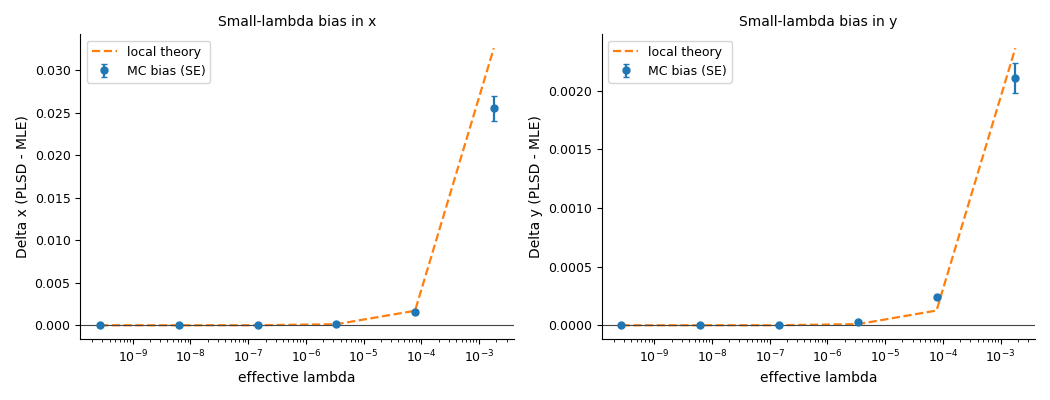}
\caption{Small-$\lambda$ linear response: heterogeneous core.}
\label{fig:smalllambda_hetero}
\end{figure}

\begin{figure}
\centering
\includegraphics[width=0.95\columnwidth]{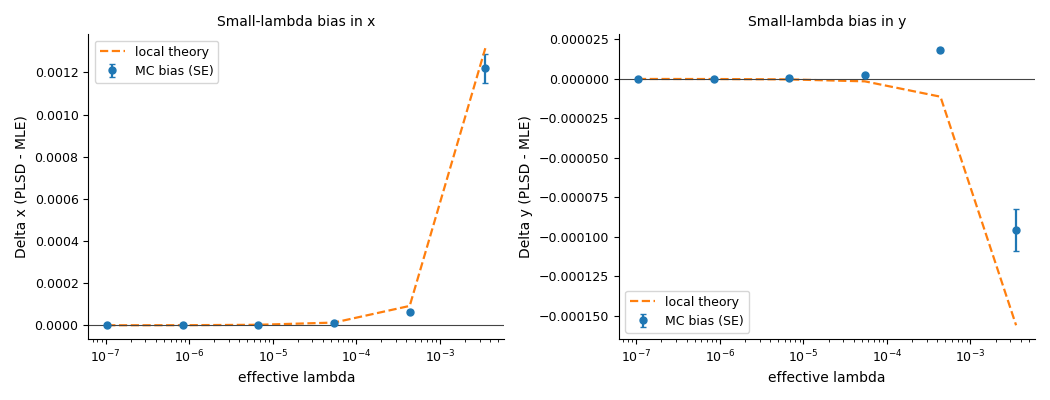}
\caption{Small-$\lambda$ linear response: triadic closure.}
\label{fig:smalllambda_triadic}
\end{figure}

For completeness, under the empirical protocol $\sigma_M$ is frozen at $\hat\theta_0$, so:
\[
Z_M(\theta)=\frac{M_{\mathrm{obs}}-\mu_M(\theta)}{\sigma_M(\hat\theta_0)},
\qquad
\nabla Z_M(\theta)=-\frac{\nabla\mu_M(\theta)}{\sigma_M(\hat\theta_0)}
\]
and hence:
\[
g_{\mathrm{pen}}(\theta)=\nabla Q(\theta)=2Z_W(\theta)\nabla Z_W(\theta)+2Z_T(\theta)\nabla Z_T(\theta)
\]
Figures~\ref{fig:smalllambda_baseline}--\ref{fig:smalllambda_triadic} show the agreement between this first-order shift and Monte Carlo estimates across scenarios.

\section{Explicit unconditional integrals}
\label{sec:appendix_unconditional}
Under the Pareto-mark layer, define $x(z)=x_{\min}/(1-z)$ for $z\in(0,1)$, $B_x(a,b)=\int_0^x t^{a-1}(1-t)^{b-1}\,dt$, and $w(z)=(\alpha-1)(1-z)^{\alpha-2}$.
The block probabilities are:
\begin{align}
P_{CC}(y) &= 2(\alpha-1)^2\int_{0}^{1}(1-z)^{2\alpha-3}
B_{z^2}\!\left(\tfrac{1}{2},1-\alpha\right)\sigmoid\!\left(2x(z)+y\right)\,dz \\
P_{CP}(y) &= \int_{0}^{1} w(z)\,\sigmoid\!\left(x(z)+y\right)\,dz \\
P_{PP}(y) &= \sigmoid(y)
\end{align}
The wedge kernels are (with $z_0$ the shared $C$ node):
\begin{align}
P^{\wedge}_{CCC}(y) &= \int_{[0,1]^3}\!\!\Bigl[\prod_{i=0}^{2} w(z_i)\Bigr]
\prod_{i=1}^{2}\sigmoid\!\left(x(z_i)+x(z_0)+y\right)\,dz_0\,dz_1\,dz_2 \\
P^{\wedge}_{CCP}(y) &= \int_{[0,1]^2}\!\!\Bigl[\prod_{i=0}^{1} w(z_i)\Bigr]
\sigmoid\!\left(x(z_0)+y\right)\sigmoid\!\left(x(z_1)+x(z_0)+y\right)\,dz_0\,dz_1 \\
P^{\wedge}_{CPP}(y) &= \int_{0}^{1} w(z)\,\sigmoid\!\left(x(z)+y\right)^2\,dz \\
P^{\wedge}_{PPP}(y) &= \sigmoid(y)^2
\end{align}
The triangle kernels are:
\begin{align}
P^{\triangle}_{CCC}(y) &= \int_{[0,1]^3}\!\!\Bigl[\prod_{i=1}^{3} w(z_i)\Bigr]
\prod_{1\le i<j\le 3}\sigmoid\!\left(x(z_i)+x(z_j)+y\right)\,dz_1\,dz_2\,dz_3 \\
P^{\triangle}_{CCP}(y) &= \int_{[0,1]^2}\!\!\Bigl[\prod_{i=1}^{2} w(z_i)\Bigr]
\sigmoid\!\left(x(z_1)+x(z_2)+y\right)\prod_{i=1}^{2}\sigmoid\!\left(x(z_i)+y\right)\,dz_1\,dz_2 \\
P^{\triangle}_{CPP}(y) &= \sigmoid(y)\int_{0}^{1} w(z)\,\sigmoid\!\left(x(z)+y\right)^2\,dz \\
P^{\triangle}_{PPP}(y) &= \sigmoid(y)^3
\end{align}

\clearpage
\bibliographystyle{elsarticle-num}
\bibliography{refs}

\end{document}